\documentclass[aps,prx,twocolumn,notitlepage]{revtex4-2}

\usepackage[utf8]{inputenc}
\usepackage{graphicx}
\usepackage{bm}
\usepackage{amsmath,amssymb,amsfonts,amsthm,mathtools}
\usepackage{braket}
\usepackage{booktabs}
\usepackage{xcolor}
\definecolor{darkgreen}{RGB}{0, 100, 0}
\usepackage{hyperref}
\usepackage{algorithm}
\usepackage{algorithmicx}
\usepackage{algpseudocode}
\begin{document}

\title{Non-Hermitian Quantum Adiabatic Algorithm}

\author{Zi-Bo Jin}
\affiliation{International Center for Quantum Materials, School of Physics, Peking University, Beijing, 100871, China}

\author{Yi Zhang}
\email{frankzhangyi@pku.edu.cn}
\affiliation{International Center for Quantum Materials, School of Physics, Peking University, Beijing, 100871, China}

\date{July 16, 2026}

\begin{abstract}
Non-Hermitian systems offer new opportunities for quantum optimization and computation. Here, we show that non-Hermitian quantum adiabatic algorithms require not only a real, gapped spectrum, but also a stable pseudospectrum. We propose a novel framework by mapping non-unitary quantum circuits to local Hamiltonian paths, thereby preserving their optimization advantages and shallow depth. While a direct non-Hermitian extension of the Feynman-Kitaev construction suffers severe pseudospectral instability, our history-decoupled construction yields both a controlled pseudospectrum and a real, gapped spectrum. Using the CK benchmark family of maximum independent set problems, we demonstrate polynomial-evolution-time non-Hermitian adiabatic computation that remains robust against perturbations. We further discuss a feasible optical implementation using coupled waveguides with an auxiliary lossy channel. Our work establishes pseudospectral stability, alongside real, gapped spectra, as key principles and a practical route for non-Hermitian quantum adiabatic computation. 
\end{abstract}

\maketitle

\section{Introduction}
\label{sec:introduction}

Quantum adiabatic algorithms (QAA) provide a distinct paradigm for quantum computation from the conventional quantum circuit (QC) paradigm \cite{BornFock1928, Kato1950, Simon1983Holonomy, KadowakiNishimori1998, Farhi2000, Farhi2001, AlbashLidar2018}. The basic idea is to encode the problem of interest into the ground state of a target Hamiltonian and, starting from a straightforward initialization, evolve sufficiently slowly along a Hamiltonian path toward it. If the energy gap between the ground state and the first excited state remains finite, the quantum adiabatic theorem guarantees that the system approximately follows the instantaneous ground state, allowing the desired solution to be read out at the end of the process. This formalism provides a Hamiltonian-based approach to various characteristic problems, such as classical optimization and quantum simulation; its efficiency is generally controlled by the minimum gap along the adiabatic path; however, finding a suitable path itself is often a nontrivial task, especially for hard problems  \cite{AminChoi2009, YoungKnyshSmelyanskiy2010, AltshulerKroviRoland2010, Choi2010, Pichler2018RydbergMIS, Ebadi2022RydbergMIS, Nguyen2023RydbergArbitraryConnectivity, Bombieri2025RydbergAdiabatic}. 

Interestingly, the Feynman-Kitaev (FK) history-state construction provides a direct connection between QC and QAA \cite{Feynman1986QuantumMechanicalComputers, KitaevShenVyalyi2002, KempeKitaevRegev2006, Aharonov2008, YuHuangWu2018ExactEquivalence}: given a QC composed of a sequence of unitary gates, whose discrete time steps are labeled by a separate clock register, the QC process is encoded as the ground state of a local, Hermitian propagation Hamiltonian. Alongside a viable route for searching viable adiabatic paths, such mapping also establishes polynomial equivalence between Hermitian QAA and unitary QC in terms of computational power \cite{Aharonov2008, KempeKitaevRegev2006, YuHuangWu2018ExactEquivalence}. 

Recently, there has been growing interest and attention in the physics of non-Hermitian systems, arising from open systems \cite{DalibardCastinMolmer1992, AshidaGongUeda2020, Minganti2020HybridLiouvillian, open1, open2, open3, open4, open5, open6, open7, open8}, dynamical systems \cite{Doppler2016DynamicallyEncirclingEP, XuMasonJiangHarris2016, AshidaGongUeda2020, Liang2022DynamicSignaturesNHSE, optical3, Xue2024, li2024obser, PhysRevB.108.214308}, systems with gain and loss \cite{BenderBoettcher1998, ElGanainy2018, MiriAlu2019, st2022, Lapp2019}, post-selection \cite{Aaronson2005PostBQP, Minganti2020HybridLiouvillian, Zhan2017NonunitaryQW, PhysRevB.109.174205}, etc., and their realizations in mechanical \cite{XuMasonJiangHarris2016, brandenbourger2019, ananya2020, chen2021, wang2022morphing, wang2023exp, li2024obser}, acoustic \cite{Christensen2016PTPhononic, ZhangYangGe2021AcousticNHSE, Zhang2021HigherOrderNHSE, ding2016, yves2017, ding2018, tang2020, puri2021, tang2021, tang2022, zhang2023exp, gu2021controlling, Wen2022, fan2023reconf, Huang2024}, optical \cite{Guo2009PT, Ruter2010PT, Regensburger2012PTLattices, MiriAlu2019, Xiao2020NHBBC, Xiao2021NonBlochPT, optical2, optical3, Xue2024, Reisenbauer2024}, and electrical platforms \cite{Helbig2020TopolectricalBBC, Liu2021ElectricalNHSE, Zou2021TopolectricalSkin, Zhu2023TopolectricalChirality, circuit1, circuit2, circuit3, circuit5}. These non-Hermitian systems can exhibit various novel phenomena, including the non-Hermitian skin effect \cite{YaoWang2018, OkumaKawabataShiozakiSato2020, ZhangYangFang2020WindingSkin, YangZhangFangHu2020aGBZ, FuZhang2023HybridSkin, ZhangZhangLuChen2022NHSEReview, YokomizoMurakami2019, lee2019ho, okugawa2020, kawabatahigher, fu2021, st2022, hu2023nonhermitianbandtheorydimensions, xiong2024nonhermitianskineffectarbitrary, liu2024localized}, novel topological phases \cite{Gong2018Topological, Kawabata2019Symmetry, BergholtzBudichKunst2021, Wang2021ComplexEnergyBraiding, RuiZhengWang2022NHSpatialSymmetries, shen2018, kunst2018bi, hu2021knot, hu2022knot, li2022braiding, guo2023knot, nakamura2024, verma2024}, non-Bloch band structures \cite{YaoWang2018, YaoSongWang2018, YokomizoMurakami2019, LeeThomale2019, YangZhangFangHu2020aGBZ, WangHYSongWang2024Amoeba, FuZhang2023Anatomy, FuZhang2024Braiding, ChenZhang2025Recurrence, hu2023nonhermitianbandtheorydimensions, yokomizo2023nonbloch, xiong2024nonhermitianskineffectarbitrary, verma2024}, unconventional criticality \cite{AshidaGongUeda2020, LiLeeGong2020CriticalNHSE, Xiao2021NHKZM, optical3, longhi2019prl, longhi2019, longhi202101, liu2020pt, liu2021}, and nontrivial dynamics \cite{Doppler2016DynamicallyEncirclingEP, XuMasonJiangHarris2016, Minganti2020HybridLiouvillian, Liang2022DynamicSignaturesNHSE, optical3, Xue2024, li2024obser, PhysRevB.108.214308, orito2022en, PhysRevB.110.035113}. Notably, unlike their Hermitian counterparts, non-Hermitian Hamiltonians generally possess complex spectra. Nevertheless, completely real spectra may still hold in several non-Hermitian scenarios \cite{ChenZhang2023RealSpectra, Zhang2022RealSpectraNoSymmetry, YangLiFuWangZhang2024ComplexSemiclassical}, including systems with $\mathcal{PT}$ symmetry \cite{BenderBoettcher1998, Bender2007, Mostafazadeh2002PseudoHermiticityII, ElGanainy2018, MiriAlu2019, Xiao2021NonBlochPT, Guo2009PT, chong2011pt, optical3, yves2017} or links to Hermitian Hamiltonians by similarity transformations \cite{ScholtzGeyerHahne1992, Mostafazadeh2002PseudoHermiticityII, AshidaGongUeda2020, ChenZhang2023RealSpectra}. 

Correspondingly, there have been considerable developments in non-Hermitian quantum computation. With non-unitary gates capable of amplification and filtering, non-unitary QCs are not only more compact, but also more straightforwardly obtainable \cite{ZhangWu2026}. Indeed, for the maximum independent set (MIS), a typical NP-complete optimization problem \cite{Karp1972, GareyJohnson1979, TarjanTrojanowski1977}, non-unitary QCs yield a simple derivation and significant amplification of the target solution's weight within polynomial depth \cite{ZhangWu2026}. In contrast, unitary QCs, e.g., following the Grover algorithm \cite{Grover1996, BennettBernsteinBrassardVazirani1997, Zalka1999}, are typically bogged down by exponential complexity. However, such exceptional advantages turn out costly: the associated success probability and resources for realizing non-unitary QCs - post-selection, gain normalization, etc. - can be formidable \cite{Aaronson2005PostBQP, Bocharov2015RUS, Silva2023FragmentedQITE, BarchLidar2025, ZhangWu2026, PhysRevB.109.174205}, limiting their physical implementations and applications. On the other hand, as QAA may resort to additional experimental implementations \cite{Johnson2011QuantumAnnealing, Boixo2014Evidence, Pudenz2014ErrorCorrectedQA, Saffman2010RydbergQI, Browaeys2020RydbergReview, WuYou2021RydbergReview, Pichler2018RydbergMIS, Ebadi2022RydbergMIS, Nguyen2023RydbergArbitraryConnectivity, Bombieri2025RydbergAdiabatic, ZhaoYouWilczekWu2025HamiltonianMIS, Shen2023YangLeeRydberg, WuYou2024DissipativeTimeCrystal, Chen2025CollectiveDissipationRydberg, Zhang2025NHManyBodyRydberg}, it may offer an alternative route to realizing non-Hermitian algorithms through a distinct continuous-time Hamiltonian implementation that may reduce, or in suitable realizations avoid, the overhead associated with repeated post-selection or gain normalization \cite{Aaronson2005PostBQP, Bocharov2015RUS, Silva2023FragmentedQITE, ZhangWu2026, BarchLidar2025}. Moreover, the finite spectral gap in QAA can provide passive protection against certain noise and control errors, making the evolution potentially more stable than a long gate sequence and alleviating the need for active error correction, magic-state distillation, and frequent intermediate measurements \cite{ChildsFarhiPreskill2002, Jordan2006, Lidar2008, YoungSarovarBlumeKohout2013, Fowler2012SurfaceCode, BravyiKitaev2005Magic, CampbellTerhalVuillot2017, GidneyFowler2019MagicFactories}; this prepare-evolve-measure structure is also favorable for near-term scaling of physical qubits \cite{Preskill2018NISQ, Johnson2011QuantumAnnealing, Boixo2014Evidence, Ebadi2022RydbergMIS}. Therefore, it is natural to study non-Hermitian quantum computation from an adiabatic-computation perspective. 

Therefore, in this work, we propose and demonstrate a novel approach to non-Hermitian QAA via mapping from non-unitary QC, whose gate sequence is converted into a local Hamiltonian path that inherits its optimization and depth advantages. Indeed, a rigorous adiabatic theorem and thus QAA still exists for non-Hermitian Hamiltonians with real spectra \cite{HuangLee2026}. However, a real, gapped spectrum alone is actually insufficient to guarantee a non-Hermitian Hamiltonian path. Unlike their Hermitian counterparts, a small perturbation to non-Hermitian Hamiltonians may result in a much larger spectral drift \cite{TrefethenEmbree2005, AshidaGongUeda2020, BergholtzBudichKunst2021, OkumaSato2021Boundary, Longhi2022SelfAcceleration, longhi202102, circuit5, hui2019, claes2021, ronika2022, kokkinakis2024anderson}. For example, the Hatano-Nelson model \cite{HatanoNelson1996, Gong2018Topological, YaoWang2018, Hatano1998, PhysRevResearch.6.L012061, PhysRevB.108.214308} - a 1D nonreciprocal chain - possesses a real spectrum under open boundary conditions; however, its properties, including its spectrum and eigenstates, are extremely sensitive to boundary perturbations \cite{TrefethenEmbree2005, OkumaKawabataShiozakiSato2020, FuZhang2023HybridSkin, HuFuZhang2025Delocalization}, i.e., the non-Hermitian skin effect \cite{YaoWang2018, OkumaKawabataShiozakiSato2020, ZhangYangFang2020WindingSkin, FuZhang2023HybridSkin, ZhangZhangLuChen2022NHSEReview, hui2019, longhi202102, claes2021, OkumaSato2021Boundary, ronika2022, circuit5, kokkinakis2024anderson}. A natural tool for characterizing this sensitivity, e.g., in the presence of perturbation, noise, control errors, etc., is the pseudospectrum \cite{TrefethenEmbree2005, OkumaSato2021Boundary, AshidaGongUeda2020, longhi202102, circuit5}. Therefore, to implement non-Hermitian QAA, we need to ensure that the Hamiltonian path has real spectra with both controllable gaps and pseudospectral stability. 

Unfortunately, a direct non-Hermitian generalization of the FK construction (FK QAA), previously used in Hermitian QC for QAA mapping, leads to an unstable pseudospectrum despite a well-behaved real, polynomially gapped spectrum; consequently, small errors accumulate and grow exponentially as the process progresses. Instead, we introduce a history-decoupled (HD) mapping, dubbed HD QAA, for non-Hermitian quantum adiabatic computations. The QC history is fully encoded in the quantum state alone, while the Hamiltonian concerns only the respective local non-unitary gate without accumulation, yielding a controlled pseudospectrum along with a real, gapped spectrum. In particular, we study the CK benchmark graph family \cite{Choi2010, Choi2011AvoidFOPT} of MIS problems ($n$ vertices) for demonstration purposes. We show that solution QCs with simple, diagonal non-unitary gates and polynomial depth [$L=\Theta(n^3)$] can be conveniently constructed. Subsequently, we present controlled non-Hermitian QAA within our HD framework, achieving polynomial wall-clock time and pseudospectral stability. Indeed, while the non-Hermitian QAA following non-unitary QC and FK construction performs satisfactorily in the noiseless case yet resists poorly against perturbations, the HD QAA exhibits remarkable success and robustness against errors; in contrast, a Hermitian QAA (HM QAA) will naturally possess a stable pseudospectrum, yet its underlying unitary QC is exponentially costly to find or in depth (QAA time); see Table~\ref{tab:qaa_comparison} for a summary. Finally, we discuss a concrete optical implementation of our non-Hermitian QAA and HD construction, effectively realized using coupled waveguides with an auxiliary lossy channel \cite{Ruter2010PT, Regensburger2012PTLattices, Zhan2017NonunitaryQW, Xiao2020NHBBC, Xiao2021NonBlochPT}.

\begin{table}[t]
\centering
\caption{Comparison of the three QAA constructions considered in this work. The HD QAA displays advantages in both efficiency and stability. }
\label{tab:qaa_comparison}
\footnotesize
\setlength{\tabcolsep}{3pt}
\renewcommand{\arraystretch}{1.15}
\begin{tabular}{lccc}
\hline
Scheme & QC/Path & Adiabatic time & Pseudospectrum \\
\hline
HM QAA & Hard & $\exp(n)$ & Stable \\
FK QAA & Easy & $\mathrm{poly}(n)$ & Unstable \\
HD QAA & Easy & $\mathrm{poly}(n)$ & Stable \\
\hline
\end{tabular}
\end{table}

The rest of this paper is organized as follows. In Sec.~\ref {sec:nh_qaa_framework}, we establish the fundamental theoretical framework of non-Hermitian QAA. We review the conventional QAA and the FK construction, the real-spectrum and pseudospectral-stability conditions for non-Hermitian QAA, and reveal the consequences of pseudospectral instability arising from non-unitary QC and FK constructions. In Sec.~\ref {sec:history_decoupled_mapping}, we introduce the HD mapping and prove that it has a real, gapped spectrum and is pseudospectrally stable. In Sec.~\ref {sec:nh_mis_algorithm}, we apply this HD QAA framework to the MIS problems, create practical non-unitary QCs and legitimate non-Hermitian QAA Hamiltonian paths, whose efficiency and stability are verified by numerical experiments, along with additional MIS benchmarks, including detailed spectral and pseudospectral results and analysis in the Appendix. We also present an implementation scheme based on coupled optical waveguides. Finally, we summarize our results and conclusions, as well as future prospects in Rydberg experiments, etc., in Sec.~\ref{sec:conclusion}. Additional derivations and details of the numerical experiments are provided in the Appendices.

\section{Non-Hermitian Quantum Adiabatic Algorithm} \label{sec:nh_qaa_framework}

In this section, we review the basic theoretical framework for non-Hermitian QAA. We start with the concepts of the QAA and the Feynman-Kitaev history-state construction - encoding a QC into the evolution of a Hamiltonian - in the usual Hermitian background. 

Then, we discuss the prerequisites for non-Hermitian QAA: the Hamiltonian on the QAA path should have a real spectrum, its ground state should be kept away from excited states by a finite energy gap, and this energy gap must remain stable under perturbations. Such stability, i.e., the spectral response to perturbations, is characterized by the pseudospectrum of the non-Hermitian Hamiltonian. 

In particular, we consider the non-Hermitian QAA following the FK construction as an example and show that, despite its satisfactorily real and gapped spectrum, the resulting pseudospectrum becomes highly unstable.

\subsection{From Quantum Circuits to Quantum Adiabatic Algorithms} \label{subsec:from_circuit_to_qaa}

\begin{figure}
    \centering
    \includegraphics[width=0.98\linewidth]{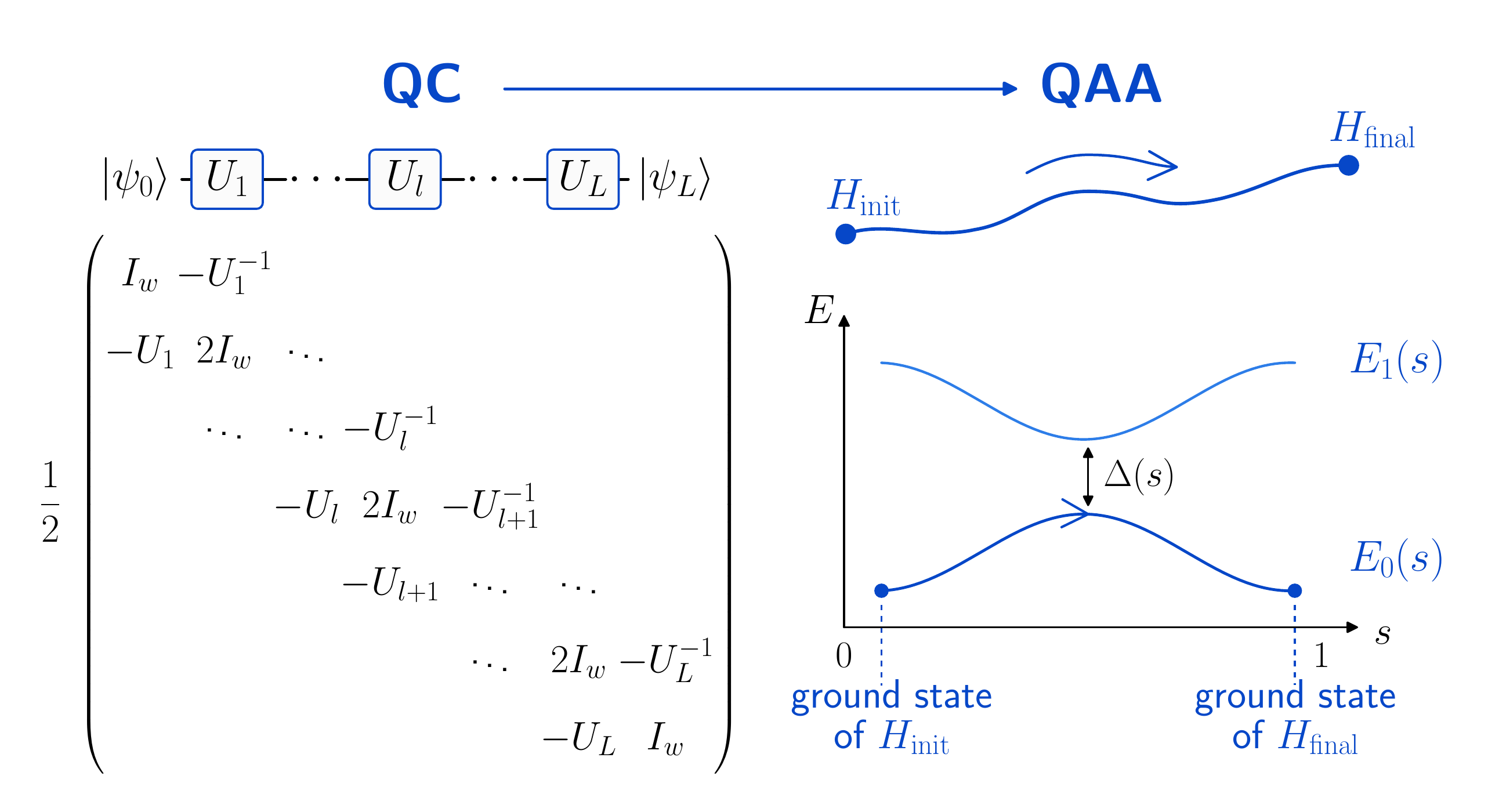}
    \caption{A schematic illustration of the mapping from QC to QAA, where the gate sequence of the QC (top left) is converted to a continuous Hamiltonian path $H(s)$ (right) from $H_\mathrm{init}$ to $H_{\mathrm{final}}$ (lower left). The spectrum remains gapped $\Delta(s)=E_1(s)-E_0(s)$, thereby protecting the adiabatic evolution of the ground state, which, upon reaching $H_{\mathrm{final}}$, yields the target solution. }
    \label{fig:qc_to_qaa_schematic}
\end{figure}

The QAA provides a quantum-computation paradigm, sometimes dubbed `analog', distinctive from the widespread `digital' QC protocol, as illustrated in the right panel of
Fig.~\ref{fig:qc_to_qaa_schematic}. For QAA, one considers a parametrized path in the Hamiltonian space: 
\begin{equation}
H(s)=(1-s)H_{\mathrm{init}}+sH_{\mathrm{final}},
\end{equation}
where $s=t/T \in[0, 1]$ is a dimensionless parameter on progress across the total evolution time $T$. $H(0)=H_{\mathrm{init}}$ is the initial Hamiltonian, whose setup and ground state are typically simple to prepare, while $H(1)=H_{\mathrm{final}}$ is the target Hamiltonian, whose ground state encodes the solution to the problem of interest. According to the quantum adiabatic theorem \cite{BornFock1928, Kato1950, Simon1983Holonomy, AlbashLidar2018}, if the instantaneous gap between the ground state and the excited states: 
\begin{equation}
\Delta(s)=E_1(s)-E_0(s),
\end{equation}
remains finite throughout the evolution, namely $\min_s \Delta(s)>0$, then, given a sufficiently slow evolution: 
\begin{equation}
T \gg \max\nolimits_{s} \frac{\| \partial_s H(s)\|}{\Delta^2(s)},
\end{equation}
the system evolves adiabatically, following the ground state at each time step, and eventually reaches the ground state of $H_{\mathrm{final}}$ at the end of the evolution, thereby obtaining the desired solution. 

However, for genuinely hard problems, following such QAA's adiabatic paths, it is customary for the Hermitian Hamiltonian to encounter gaps $\Delta(s)$ that are exponentially small in system sizes, thereby leading to exponentially long evolution times and squandering the quantum advantage. Even if well-designed Hamiltonian paths may avoid such exponentially closing gaps, finding these optimized paths is itself a highly nontrivial quantum many-body optimization problem. 

Alternatively, one can guarantee a valid QAA Hamiltonian path via the Feynman-Kitaev (FK) construction based on a working QC that solves the same problem of interest \cite{Aharonov2008, YuHuangWu2018ExactEquivalence}. A QC consists of a sequence of unitary gates, $U_1, U_2, \ldots, U_L$, as depicted in the top left panel of Fig.~\ref{fig:qc_to_qaa_schematic}, which transforms a trivial initial state $|\psi_0 \rangle$ to a final state hosting the target solution $ |\psi_L \rangle = U_L U_{L-1} \ldots U_2 U_1 |\psi_0 \rangle$. $U_l^{-1}=U_l^{\dagger}$ encodes information of the QC and, in turn, the original problem, and $L$ is the circuit depth. The FK construction maps such a step-by-step QC into a time-dependent Hamiltonian path: 
\begin{equation}
    H(s) = s H_\mathrm{FK} + (1-s) H_{init},
    \label{eq:HFK_QAA}
\end{equation}
where: 
\begin{equation}
H_\mathrm{init}=I_w\otimes \sum_{l=1}^{L}|l\rangle\langle l|,
\end{equation}
and:
\begin{equation}
H_{\mathrm{FK}}=\sum_{l=1}^{L}h_l, \label{eq:HFK}
\end{equation}
\begin{eqnarray}
h_l&=&\frac{1}{2}\Big(
I_w\otimes\ket{l-1}\bra{l-1}
+I_w\otimes\ket{l}\bra{l} \nonumber  \\
& &-U_l\otimes\ket{l}\bra{l-1}
-U_l^{-1}\otimes\ket{l-1}\bra{l}
\Big), \label{eq:hqaa2}
\end{eqnarray}
as illustrated in the lower left panel in Fig.~\ref{fig:qc_to_qaa_schematic}. Here, the Hamiltonian $H(s)$ acts on the composite space $\mathcal H_w \otimes \mathcal H_c$, where $\mathcal H_w$ is the $2^n$-dimensional work space consists of the $n$ qubits of the target problem, and $\mathcal H_c=\mathrm{span}\{\ket{0}, \ket{1}, \ldots, \ket{L}\}$ is a $(L+1)$-dimensional clock space that labels the logical steps in correspondence to the QC. $I_w$ denotes the identity operator in the work space. 

Let us define the cumulative circuit after the $l$-th step as $W_l=U_lU_{l-1}\cdots U_1$, with the convention $W_0=I$. As $s$ gradually increases from 0, $H_{\mathrm{FK}}$ introduces off-diagonal terms coupling the bases $|\psi_l\rangle$ and $|\psi_{l-1}\rangle$, where $|\psi_l\rangle=U_l|\psi_{l-1}\rangle$ and $|\psi_{l-1}\rangle=U_l^{-1}|\psi_l\rangle$. Therefore, given an initial state $\ket{\psi_0}$ ($\ket{\Psi_0}=\ket{\psi_0}\otimes\ket{0}$ in the composite space), after successive adiabatic evolution, we obtain the final ground state: 
\begin{equation}
\ket{\Psi_L} = \frac{1}{\sqrt{L+1}}\sum_{l=0}^{L}W_l\ket{\psi_0}\otimes\ket{l},
\end{equation}
see further details in Ref.~\cite{Aharonov2008,KempeKitaevRegev2006} and the Appendix. Importantly, the minimum energy gap $\Delta_s \sim \Theta(L^{-2})$ between the ground state and excited states across the entire Hamiltonian path decreases only polynomially with the step size $L$. Therefore, we can ensure the algorithm's adiabaticity for an elapsed time $T$ that scales polynomially with $L$. The target solution, namely $W_L\ket{\psi_0}$ in the workspace $\mathcal H_w$, can be extracted by projecting the clock state onto $\ket{L}$, e.g., following a measurement with probability $1/(L+1)$, at the end of the evolution $s=1$ and $t=T$. 

Thus, such an FK construction provides a feasible Hamiltonian path through a mapping from QC. In practice, however, designing a Hermitian QC with a provable advantage over classical algorithms is still highly challenging. In contrast, non-Hermitian or non-unitary QCs are often more straightforward to design, owing to mechanisms such as amplification and filtering. Gate-based implementations of non-unitary operations often rely on repeated post-selection or gain normalization and suffer from accumulated errors and substantial fault-tolerant costs as the circuit depth increases \cite{Aaronson2005PostBQP, Bocharov2015RUS, Silva2023FragmentedQITE, ZhangWu2026}. By contrast, mapping them to QAA offers an alternative continuous-time physical realization that may reduce such reliance, while its gapped prepare-evolve-measure structure may enhance robustness and favor near-term scalability \cite{ChildsFarhiPreskill2002, Lidar2008, Preskill2018NISQ, Ebadi2022RydbergMIS}. In addition, the extra degrees of freedom provided by non-Hermitian QAA may partially alleviate path constraints and thus reduce the design difficulties in Hermitian QAA. Subsequently, we will extend our consideration of $H(s)$ paths to the non-Hermitian Hamiltonian space.

\subsection{The real-spectrum condition for non-Hermitian QAA} \label{subsec:real_spectrum_condition}

Compared with Hermitian QAAs, where the spectrum of $H(s)$ is inherently real-valued, a non-Hermitian $H(s)$ allows a complex spectrum, and thus, we require our non-Hermitian QAA to possess a real spectrum as a prerequisite, which allows an unambiguous definition of an energy gap that isolates the ground state and protects the adiabaticity. In addition, the imaginary part of a complex eigenenergy corresponds to a dynamical gain or loss. After long-time evolution, the final state will be exponentially dominated by the eigenstate with the largest gain (smallest decay), rather than the target ground state by the adiabatic principle. Indeed, recent studies have shown that a rigorous adiabatic theorem can be formulated for quantum evolution of non-Hermitian Hamiltonians with real spectra~\cite{HuangLee2026}. 

A real spectrum may result from several different mechanisms, e.g., $\mathcal{PT}$-symmetry \cite{BenderBoettcher1998, Mostafazadeh2002PseudoHermiticityII, Guo2009PT, chong2011pt}, or, even in the absence of symmetry, certain criteria of the generalized Brillouin zone \cite{ChenZhang2023RealSpectra}. More universally, a necessary and sufficient criterion for real spectra exists for finite-dimensional diagonalizable non-Hermitian Hamiltonians - it can be transformed to a Hermitian Hamiltonian $H_0=H_0^\dagger$:  
\begin{equation}
H=SH_0S^{-1}, 
\label{eq:similarity_transform}
\end{equation}
by an invertible similarity transformation $S$: 
\begin{equation}
S=\sum_n |r_n\rangle\langle n|, \quad S^{-1}=\sum_n |n\rangle\langle l_n|,
\end{equation}
where $\{|n\rangle\}$ is an orthonormal basis, and $|r_n\rangle$ ($\langle l_n|$) are the right (left) eigenstates of $H$ satisfying the biorthonormal condition $\langle l_m|r_n\rangle=\delta_{mn}$. Clearly, $H_0$ and $H$ share an identical, real spectrum.

\subsection{The stable-pseudospectrum condition for non-Hermitian QAA} \label{subsec:pseudospectrum_stability}

In addition to the real-spectrum condition, a non-Hermitian QAA must also take spectral stability into account. Eq.~\eqref{eq:similarity_transform} provides a real-spectrum condition that the non-Hermitian Hamiltonian should be isospectral to a Hermitian Hamiltonian through a similarity transformation $S$; however, this does not guarantee that these eigenvalues within the spectrum remain stable under perturbations. Indeed, a non-Hermitian Hamiltonian $H$ is generally non-normal, and a non-unitary similarity transformation $S$ may stretch, compress, and tilt the orthogonal eigenstates of the corresponding Hermitian Hamiltonian $H_0$, making them non-orthogonal and sometimes even nearly parallel. The distinctions between these eigenstates rely on their linearly independent components, which, when small, become sensitive to small perturbations, leading to frequent mixing of eigenvalues and an unstable spectrum. 

This geometric instability of the eigenbasis can be characterized by the condition number \cite{TrefethenEmbree2005} of the similarity transformation matrix: 
\begin{equation}
\kappa(S) = \|S\|\,\|S^{-1}\| = \frac{\sigma_{\max}(S)}{\sigma_{\min}(S)}\ge 1,
\label{eq:condition_number}
\end{equation}
where $\sigma_{\max}(S)$ and $\sigma_{\min}(S)$ are the largest and smallest singular values of $S$, respectively. Physically, $\kappa(S)$ measures the ratio between maximal stretching and compression of $S$ to the state space. If $\kappa(S)\gg 1$, certain eigenstates are strongly stretched and nearly linearly dependent. The extreme case is $\kappa(S)\to\infty$ approaching an exceptional point, where eigenstates coalesce asymptotically. 

More specifically, according to the Bauer-Fike estimate \cite{BauerFike1960, TrefethenEmbree2005}, under a small perturbation:
\begin{equation}
H(s)\rightarrow H(s)+\delta H{(s)}, 
\end{equation}
the distance of any perturbed eigenvalue $E'_n$ to the original spectrum is bounded by: 
\begin{equation}
\min_m |E'_n-E_m| \leq \kappa(S)\|\delta H{(s)}\|.
\end{equation}
where $\|\cdot\|$ denotes the matrix two-norm, i.e., the operator norm induced by the Euclidean vector norm, or equivalently, the largest singular value. Therefore, given a large $\kappa(S)$, a small perturbation may be strongly amplified, inducing a much larger spectral response and thereby forfeiting the spectral gap and the adiabatic condition: 
\begin{equation}
\kappa(S)\|\delta H{(s)}\|\sim \Delta(s) ,
\end{equation}
where $\Delta(s)$ is the instantaneous spectral gap protecting adiabaticity. 

\begin{figure*}
    \centering
    \includegraphics[width=0.9\linewidth]{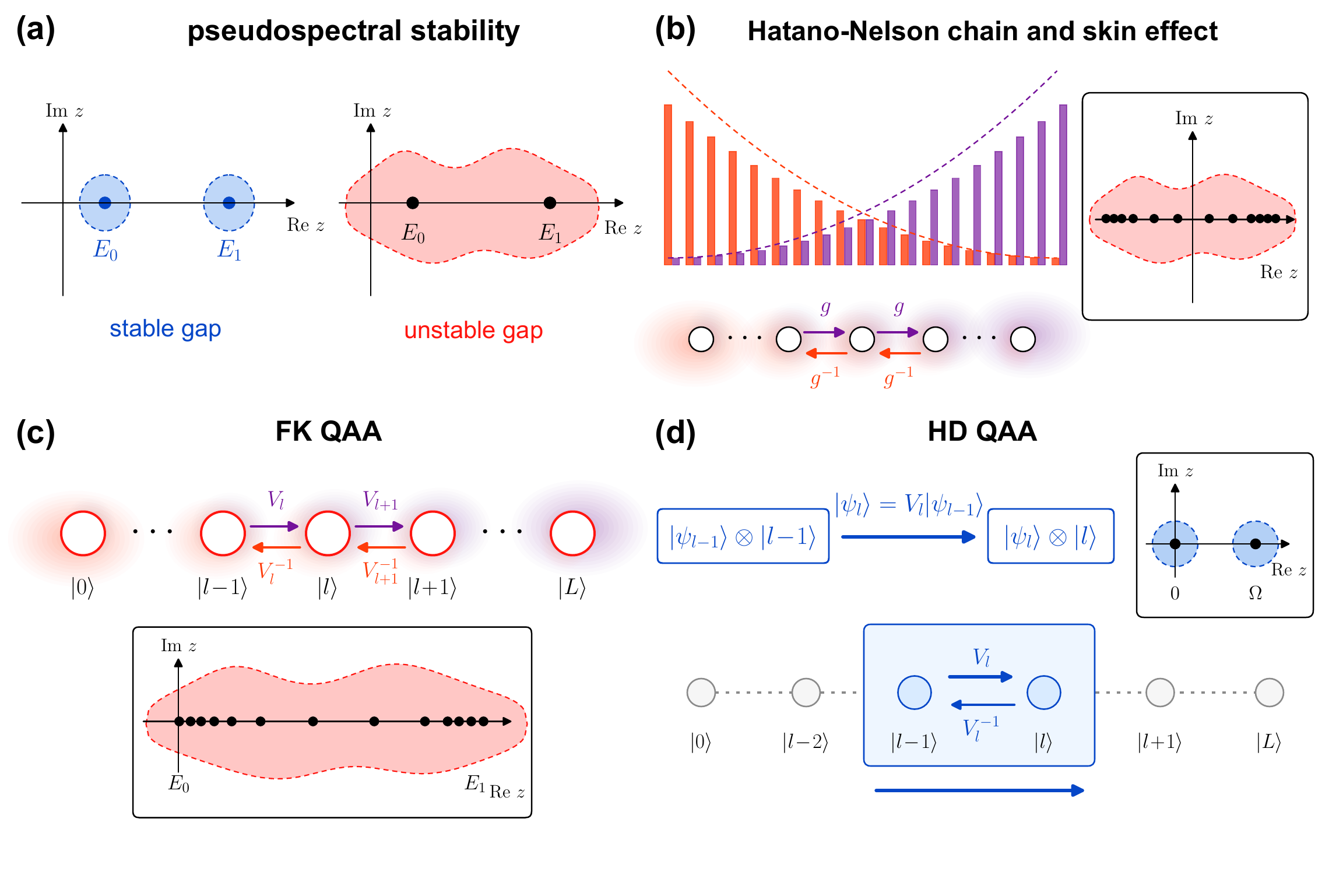}
    \caption{(a) A gap with an isolated (merged) pseudospectrum is stable (unstable) under perturbations. (b) The non-Hermitian Hatano-Nelson model exhibits a highly unstable pseudospectrum, thus the non-Hermitian skin effect and sensitivity to perturbations such as boundary conditions. (c) The FK QAA's effective model resembles a non-reciprocal Hatano-Nelson chain, in which the non-normality of the similarity transformation accumulates, leading to an unstable pseudospectrum. (d) In comparison, the HD QAA converts non-unitary gates $V_l$ and $V_l^{-1}$ and establishes non-reciprocal coupling locally, thereby ensuring overall pseudospectral stability and, thus, adiabaticity under perturbations. }
    \label{fig:schematic_psedospectrum}
\end{figure*}

Quantitatively, a natural tool for characterizing this sensitivity to perturbations is the pseudospectrum \cite{TrefethenEmbree2005, AshidaGongUeda2020, OkumaSato2021Boundary, longhi202102, circuit5}. For a Hamiltonian $H$ and a threshold $\varepsilon>0$, its $\varepsilon$-pseudospectrum is defined as: 
\begin{equation}
\label{eq:resolvent_pesedospectrum}
\Lambda_\varepsilon(H) = \left\{
z\in\mathbb C: 
\|(zI-H)^{-1}\|\geq \varepsilon^{-1}
\right\}.
\end{equation}
Equivalently, it equals the union of spectra following all perturbations below amplitude $\varepsilon$: 
\begin{equation}
\label{eq:perturbations_pseudospectrum}
\Lambda_\varepsilon(H) = \bigcup_{\|\Delta H\|\leq \varepsilon}
\mathrm{spec}(H+\Delta H). 
\end{equation} 
Thus, the pseudospectrum characterizes the complex region that the eigenvalues may span under small perturbations, as schematically illustrated in Fig.~\ref{fig:schematic_psedospectrum}(a). 

We can make the following theoretical estimate on the pseudospectral contour for small perturbations $\varepsilon$. Denoting the eigenvalues and projectors of a diagonalizable Hamiltonian $H$ as $E_j$ and $\Pi_j$, respectively, we can expand the resolvent as: 
\begin{equation}
	(zI-H)^{-1}=\sum_j \Pi_j/(z-E_j). 
\end{equation} 
Rewriting $z=E_j+\rho e^{i\phi}$ near $E_j$, we obtain: 
\begin{equation}
	(zI-H)^{-1}=\Pi_j/(\rho e^{i\phi})+\sum_{k\neq j}\Pi_k/(E_j-E_k+\rho e^{i\phi}). 
\end{equation}
Given a small perturbation $\varepsilon$, each pseudospectral region is dominated separately by a single pole $E_j$, i.e., $\rho\ll \min_{k\neq j}|E_j-E_k|$, we can approximate $(zI-H)^{-1}\simeq \Pi_j/(\rho e^{i\phi})$. Therefore, the pseudospectral contour $\|(zI-H)^{-1}\|=\varepsilon^{-1}$, following Eq.~\eqref{eq:resolvent_pesedospectrum}, yields a circle of radius: 
 \begin{equation}
     \rho_j\simeq\varepsilon\|\Pi_j\|. 
     \label{eq:rhoestimate}
 \end{equation} 
Above a perturbation threshold $\varepsilon>\varepsilon_c$, the pseudospectrum merges between $E_0$ and $E_1$ and closes the gap,  $\Delta(s) \sim |E_0-E_{1}| \sim \varepsilon_c (\|\Pi_0\| + \|\Pi_{1}\|)$, and the adiabatic condition is violated.

Further, for Hamiltonians satisfying Eq.~\eqref{eq:similarity_transform}, the size of $\Lambda_\varepsilon(H)$ is upper bounded by the condition number in Eq.~\eqref{eq:condition_number}: 
\begin{eqnarray}
\|(zI-H)^{-1}\| &\leq& \|S\|\, \|(zI-H_0)^{-1}\|\, \|S^{-1}\| \nonumber \\
&=& \kappa(S) \|(zI-H_0)^{-1}\|. 
\end{eqnarray}
Thus, one obtains an approximate upper-bound estimate: 
\begin{equation}
\Lambda_\varepsilon(H) \subset \left\{z\in\mathbb C: \mathrm{dist}\bigl(z,\mathrm{spec}(H_0)\bigr) \leq \kappa(S)\varepsilon\right\}, 
\label{eq:condition_number_estimate}
\end{equation}
where, for the Hermitian Hamiltonian $H_0$: 
\begin{equation}
\|(zI-H_0)^{-1}\| = \frac{1} {\mathrm{dist}\bigl(z,\mathrm{spec}(H_0)\bigr)}. 
\end{equation}
Therefore, the expansion of the pseudospectrum in the complex plane has an upper bound that is directly proportional to $\kappa(S)$, and a large $\kappa(S)$ can substantially enlarge the complex region that the eigenvalues may scatter under small perturbations, thereby reflecting strong perturbation sensitivity. 

Recent studies in non-Hermitian band theory have also shown that pseudospectral instability is not an exception limited to special models, but rather widespread in non-Hermitian systems \cite{OkumaSato2021Boundary, longhi202102, circuit5, hui2019, claes2021, ronika2022, kokkinakis2024anderson}. A typical example is the one-dimensional Hatano--Nelson chain on $L+1$ sites, indexed by $j=0, 1, \ldots, L$, with open boundary
conditions \cite{Hatano1998, PhysRevResearch.6.L012061, HatanoNelson1996}: 
\begin{equation}
H_{\mathrm{HN}} = \sum_{j=0}^{L-1} \left[ g\,|j+1\rangle\langle j| + g^{-1}|j\rangle\langle j+1| \right], 
\end{equation}
where $g>0$ characterizes the non-reciprocity between leftward and rightward hopping amplitudes. $H_{\mathrm{HN}} $ can be mapped to a Hermitian chain:
\begin{equation}
H_0 = \sum_{j=0}^{L-1} \left[ |j+1\rangle\langle j| + |j\rangle\langle j+1| \right], 
\end{equation}
through a non-unitary similarity transformation $S= \mathrm{diag} \left( 1,g,g^2,\cdots,g^{L} \right)$: $H_{\mathrm{HN}} = S H_0 S^{-1}$. Therefore, $H_{\mathrm{HN}}$ and $H_0$ have an identical, real spectrum. 

However, the Hatano-Nelson model may suffer from an unstable pseudospectrum, as its condition number grows exponentially with the system size $L$: 
\begin{equation}
\kappa(S) = \frac{\max_j |g|^{j}}{\min_j |g|^{j}} = e^{L|\ln g|}, 
\end{equation}
as long as $g\neq 1$. We note that when $g=1$, the leftward and rightward hopping amplitudes are equal, and the system reduces to a Hermitian chain; in contrast, when $g\neq 1$, the hopping amplitudes are nonreciprocal, and $H_{\mathrm{HN}}$ is non-Hermitian. Indeed, the similarity transformation $S$ multiplies the eigenstates of the Hermitian chain by a position-dependent factor $g^j$, making them accumulate near the right (left) boundary for $g>1$ ($0<g<1$) and producing the non-Hermitian skin effect, as illustrated in Fig.~\ref{fig:schematic_psedospectrum}(b). Also, small perturbations on the boundary can lead to pronounced spectral responses \cite{OkumaKawabataShiozakiSato2020}. Similar phenomena are also typically observed in other nonreciprocal chains: while a real spectrum may be present under open boundary conditions, the condition number explodes with system size, leading to substantial pseudospectral instability.

We emphasize that pseudospectral stability is paramount for non-Hermitian QAA. A finite gap is necessary to protect the ground state and its adiabatic evolution; if, however, this gap exists only in the ideal scenario and is unstable to perturbations, e.g., noise, control errors, or model imperfections, as in various non-Hermitian systems, the QAA may fail, as we demonstrate in the next subsection.

\subsection{Example: unstable pseudospectrum of non-Hermitian QAA through FK construction} \label{subsec:fk_unstable_pseudospectrum}

First, we consider in this subsection the direct non-Hermitian generalization of the FK construction. Given a non-unitary QC composed of invertible non-unitary gates $V_L, V_{L-1}, \cdots, V_1, V_l\in GL(\mathcal H_w)$, we replace the QAA Hamiltonian setup in Eqs. \eqref{eq:HFK_QAA}, \eqref{eq:HFK}, and \eqref{eq:hqaa2} as $U_l\longrightarrow V_l, U_l^{-1}\longrightarrow V_l^{-1}$: 
\begin{equation}
\begin{aligned}
h_l=\frac{1}{2}\Big(
&I_w\otimes |l-1\rangle\langle l-1|
+I_w\otimes |l\rangle\langle l|  \\
&-V_l\otimes |l\rangle\langle l-1|
-V_l^{-1}\otimes |l-1\rangle\langle l|
\Big). 
\end{aligned}
\end{equation}
Since $V_l^{-1}\neq V_l^\dagger$ is non-unitary, $h_l$, and correspondingly, $H(s)$, is generally non-Hermitian. 

Furthermore, $H(s)$ retains a real, gapped spectrum like the Hermitian case, as is clear from the similarity transformation $S = \sum_{l=0}^{L} W_l\otimes |l\rangle\langle l|$: 
\begin{equation}
H(s)= S \left[ sH_\mathrm{clock} + (1-s)H_{init}\right] S^{-1}, 
\label{eq:Hclock}
\end{equation}
where: 
\begin{equation}
H_{\mathrm{clock}}=\frac{I_w}{2}\sum_{l=1}^{L}\bigl(|l\rangle-|l-1\rangle\bigr)\bigl(\langle l|-\langle l-1|\bigr).
\end{equation}
Especially, the final output state is, up to normalization,
$|\Psi_L \rangle = \sum_{l=0}^{L} W_l|\psi_0 \rangle \otimes |l\rangle$, which incorporates the solution-bearing target state $\ket{\psi_L}=W_L\ket{\psi_0}$ in the work space. Here, $W_l=V_lV_{l-1}\cdots V_1$, and $W_0=I$ similar to the unitary case. 

Unfortunately, however, the FK construction does not guarantee a stable pseudospectrum, as schematically shown in Fig.~\ref{fig:schematic_psedospectrum}(c). As an explicit example, we consider the maximum independent set (MIS) problem, whose details and non-unitary QC solution will be discussed in Sec.~\ref{subsec:nh_mis_algorithm}. After mapping the non-unitary QC to a non-Hermitian QAA through the FK construction (FK QAA), we evaluate the pseudospectra (of the ground state and the first excited state) based on numerical calculations (Eq.~\eqref{eq:resolvent_pesedospectrum}) and theoretical estimates (Eq. ~\eqref{eq:rhoestimate}), and further determine the noise threshold $\varepsilon_c$ that leads to gap closure, as summarized in Fig.~\ref{fig.exp_two_psedospectrum}. Clearly, even though the FK construction yields a non-Hermitian QAA Hamiltonian with a real spectrum, its gap is highly sensitive to small perturbations and, from time to time, completely toppled, along with the adiabaticity that sustains quantum computation. 

The origin of such pseudospectral instability can be traced back to the FK construction's simultaneous encoding of the entire history of the non-unitary QC. After the FK construction, the effective Hamiltonian resembles a Hatano-Nelson model of length $L+1$, where the nonreciprocal hopping corresponds to $V_l$. While its non-Hermitian skin effect may enhance the weight over $|L\rangle$ and thus the solution probability, the pseudospectral instability inevitably accumulates along the entire nonreciprocal chain. Equivalently, the similarity transformation $S=\sum_{l=0}^{L}W_l\otimes |l\rangle\langle l|$ points to a condition number $\kappa(S)$ dominated by $W_l$, which accumulates the singularities from multiple $V_l$ non-unitary gates exponentially. Thus, a main conclusion of this work and the focus of the next sections is why and how to truncate such accumulation of pseudospectral instability and limit it to a single respective $V_l$ at each step.

\section{Non-Hermitian Quantum Adiabatic Algorithm through history-decoupled construction} \label{sec:history_decoupled_mapping}

Previously, we have shown that the non-Hermitian QAA, constructed via the FK method, exhibits a real, sufficiently gapped spectrum but is hampered by a highly unstable pseudospectrum. To stem such instability, in this section we introduce a history-decoupled (HD) QAA: the instantaneous Hamiltonian is responsible only for the current non-unitary gate $V_l$, whereas the history of the quantum evolution is encoded fully and only in the quantum state. 

We start with a division of the total evolution time $T$ into $L$ segments, and define: 
\begin{equation}
s(t)=Lt/T-(l-1)\in[0,1] \quad \mbox{for} \quad t\in\left[\frac{l-1}{L}T,\frac{l}{L}T\right],
\end{equation}
in the $l$-th segment. Conversely, given $l$ and $s$, we can uniquely locate the time instance $t= (l-1+s)T/L$. $t\in [0, T]$ and $l=1, 2, \cdots, L$. Following the gate sequence $V_l$ in the non-unitary QC, the non-Hermitian Hamiltonian following our HD QAA writes as: 
\begin{equation}
\label{eq:HD_Hamiltonian}
H(t)=H_l\bigl(s(t)\bigr),  \quad H_l(s)=\Omega\bigl(I-P_l(s)\bigr),
\end{equation}
where $\Omega>0$ is a parameter for the gap, and: 
\begin{equation}
P_l(s)=\sum_a \ket{r^{(l,a)}(s)} \bra{\ell^{(l,a)}(s)},
\end{equation}
is a projection operator in a dynamical bi-orthogonal basis of the composite space $\mathcal H_w \otimes \mathcal H_c$: 
\begin{eqnarray}
\ket{r^{(l,a)}(s)} &=& \cos\theta(s)\ket a\otimes\ket{l-1} + \sin\theta(s)V_l\ket a\otimes\ket l, \nonumber\\
\bra{\ell^{(l,a)}(s)} &=& \cos\theta(s)\bra a\otimes\bra{l-1} + \sin\theta(s)\bra a V_l^{-1}\otimes\bra l, \nonumber\\
& &\braket{\ell^{(l,a)}(s)|r^{(l,b)}(s)}=\delta_{ab},
\end{eqnarray} 
where $\{\ket a\}$ is an orthonormal basis of the work space $\mathcal{H}_w$, $\mathcal{H}_c = \mathrm{span}\{\ket{0}, \ket{1}, \cdots, \ket{L}\}$ is an auxiliary clock space, and $\theta(s)$ is a smooth monotone function, $\theta(0)=0$, $\theta(1)=\pi/2$, to evolve $H_l(0)$ to $H_l(1)$. Since: 
\begin{equation}
H_{l}(1)=H_{l+1}(0)=\Omega\left(I-I_w\otimes |l\rangle\langle l|\right), 
\end{equation} 
such a Hamiltonian path is continuous and smooth \footnote{We can also enforce smoothness (continuous $\dot H(t)$) at the segment boundaries by choosing an interpolation function that satisfies $\theta'(0)=\theta'(1)=0$. } both within each segment and at the connecting points between adjacent segments. 

Although both HD QAA and FK QAA introduce a clock space, the two constructions encode and execute the circuit information differently. In the FK QAA, the complete QC is written at once into a static global Hamiltonian $H_{\mathrm{FK}}$ in Eq.~\eqref{eq:HFK}. By contrast, the HD QAA Hamiltonian concerns only the $l$-th layer of the QC, $V_l$ and its inverse $V_l^{-1}$, coupling only the neighboring $|l-1\rangle$ and $|l\rangle$ sectors. In short, the $V_l$ gates enter one by one, rather than simultaneously, into a single clock-chain Hamiltonian. 

Importantly, for an initial state $|\psi_{l-1}\rangle\in\mathcal H_w$ and thus $|\psi_{l-1}\rangle\otimes |l-1\rangle$ in the composite space at the beginning of the $l$-th segment, the QAA gradually turns it into:  
\begin{equation}
|\psi_l\rangle\otimes |l\rangle, \quad |\psi_l\rangle=V_l|\psi_{l-1}\rangle, 
\end{equation} 
at the end of the segment. Repeating this process, segment by segment, we arrive at the final quantum state harboring the solution $|\psi_L\rangle = V_L \cdots V_2 V_1|\psi_0\rangle$, as directed by the original non-unitary QC. Moreover, since $P_l(s)^2=P_l(s)$ is a projection operator, $H_l(s)$ and thus $H(t)$ possess a two-level spectrum: $0$ and $\Omega$, ensuring a real and finitely gapped spectrum and the first condition for non-Hermitian QAA. These properties are similar to those of the FK QAA; however, we show next that our HD QAA also retains a stable pseudospectrum. 

We can compute the pseudospectrum of this Hamiltonian analytically following Eq.~\eqref{eq:resolvent_pesedospectrum}. Take the singular-value decomposition: 
\begin{equation}
V_l=U_l\Sigma_l R_l^\dagger, \quad \Sigma_l=\mathrm{diag} (\sigma_1,\sigma_2,\ldots,\sigma_{d_w}),    
\end{equation}
where $d_w=\dim\mathcal H_w$, we define the unitary transformation:
\begin{equation}
Q_l = R_l\otimes |l-1\rangle\langle l-1| + U_l\otimes |l\rangle\langle l| + I_w\otimes \sum_{m\neq l-1,l} |m\rangle\langle m|,
\end{equation}
which, while retaining its pseudospectrum, changes $H_l$ to the basis: 
\begin{equation}
Q_l^\dagger H_l(s)Q_l = \left[ \bigoplus_{j=1}^{d_w} H_{\sigma_j}(s) \right] \oplus \Omega I_{\mathrm{inact}}, 
\end{equation}
where $I_{\mathrm{inact}}$ acts on all inactive clock sites, and: 
\begin{equation}
H_{\sigma_j}(s)
=
\frac{\Omega}{2}
\begin{pmatrix}
    1-\cos 2\theta(s)
    &
    -\sin 2\theta(s)\,\sigma_j^{-1}
    \\
    -\sin 2\theta(s)\,\sigma_j
    &
    1+\cos 2\theta(s)
\end{pmatrix}.
\end{equation}
Therefore, we can obtain the full $\varepsilon$-pseudospectrum through the disk-shape pseudospectrum $\{z\in\mathbb C:|z-\Omega|\leq\varepsilon\}$ of the diagonal $\Omega I_{\mathrm{inact}}$, together with the pseudospectra of these decoupled $H_{\sigma_j}(s)$: 
\begin{equation}
|z(z-\Omega)|^2 = \varepsilon^2 \left[\mathrm{tr} \left( (zI_2-H_{\sigma_j})^\dagger (zI_2-H_{\sigma_j}) \right) -\varepsilon^2 \right],
\end{equation}
where:
\begin{equation}
\begin{aligned}
\mathrm{tr}
\left[ (zI_2-H_{\sigma_j})^\dagger (zI_2-H_{\sigma_j}) \right]
&=
\left|z-\frac{\Omega}{2}\left[1-\cos 2\theta(s)\right]\right|^2
\\
+
\left|z-\frac{\Omega}{2}\left[1+\cos 2\theta(s)\right]\right|^2
+{}&
\frac{\Omega^2}{4}\sin^2 2\theta(s)
\left(\sigma_j^2+\sigma_j^{-2}\right),
\end{aligned}
\end{equation}
which gives the boundary of the pseudospectrum. The $\varepsilon$-pseudospectrum is the area enclosed by the boundary satisfying $\sigma_{\min}(zI_2-H_{\sigma_j})\leq\varepsilon$; see detailed derivation in Appendix. In the end, the pseudospectrum is controlled only by the singular-value combination $\sigma_j^2+\sigma_j^{-2}$ of a local gate and does not accumulate across the entire QC. 

Alternatively, we can examine the similarity transformation that maps the current Hamiltonian to its identity-gate counterpart, Eq.~\eqref{eq:HD_Hamiltonian} with $V_l=I_w$ in $P_l(s)$: 
\begin{equation}
S_l = I_w\otimes\left(I_c-|l\rangle\langle l|\right) + V_l\otimes |l\rangle\langle l|,
\end{equation}
whose condition number controls the (upper-bound) pseudospectral thickness: 
\begin{equation}
\kappa(S_l) = \max\{1,\|V_l\|\} \max\{1,\|V_l^{-1}\|\}, 
\end{equation}
receives only contribution locally from a single non-unitary gate $V_l$ and limits its accumulation, instead of the entire depth of the QC history $W_l$ as in the FK construction. 

Physically, the HD mapping does not eliminate the non-Hermiticity; rather, it rearranges its organization in the clock space. In the FK construction, all states in the clock space are linked simultaneously; the imaginary gauge field, representing the singular-value imbalance across each link, accumulates across the entire nonreciprocal direction; as a result, the Green's function diverges as it crosses multiple links, indicating unstable responses to perturbations, state-wise or spectrum-wise. In contrast, only the neighboring states in the clock space are locally linked at any given time, truncating the gain and amplification of the imaginary gauge field and the Green's function across multiple links; instead, the past amplification and filtering effects of the non-unitary QC - the history - are fully encoded in the quantum state. It is essentially this decoupling from history that allows our non-Hermitian QAA to establish a better-behaved instantaneous Hamiltonian and limit the extent of its non-normality and pseudospectrum, as schematically illustrated in Fig.~\ref{fig:schematic_psedospectrum}(d), thus retaining the quantum-computation outcomes of the non-unitary QC while avoiding the cumulative pseudospectral instability introduced by the FK construction. 

In addition to the resulting non-Hermitian QAA's pseudospectrum stability, the HD QAA is also more efficient than the FK QAA in the ideal case. Indeed, our subsequent Hamiltonian has a constant gap $\Omega$, rather than the $\Theta(L^{-2})$ polynomial gap in the FK QAA, and largely suppresses the required evolution time to achieve adiabaticity. Besides, the HD QAA places the final state entirely in the $|L\rangle$ clock sector. In the Hermitian FK construction, this sector has a weight of $1/(L+1)$, introducing an $O(L)$ readout overhead. In the non-Hermitian case, however, the weight of the $|L\rangle$ sector can be $O(1)$ or exponentially small, depending on the presence of either the rightward or leftward NHSE following the corresponding FK construction.

\section{Realistic Non-Hermitian QAA for Maximum Independent Set} \label{sec:nh_mis_algorithm}

We have introduced the HD QAA in the previous section. In this section, we focus on the MIS problem and demonstrate that, combined with efficient non-unitary QC, a non-Hermitian QAA Hamiltonian path that is both effective and robust can be obtained through the HD construction. Further, we discuss potential experimental realizations in optical platforms.

\subsection{Non-Hermitian algorithms for the maximum independent set} \label{subsec:nh_mis_algorithm}

\begin{figure}
    \centering
    \includegraphics[width=0.9\linewidth]{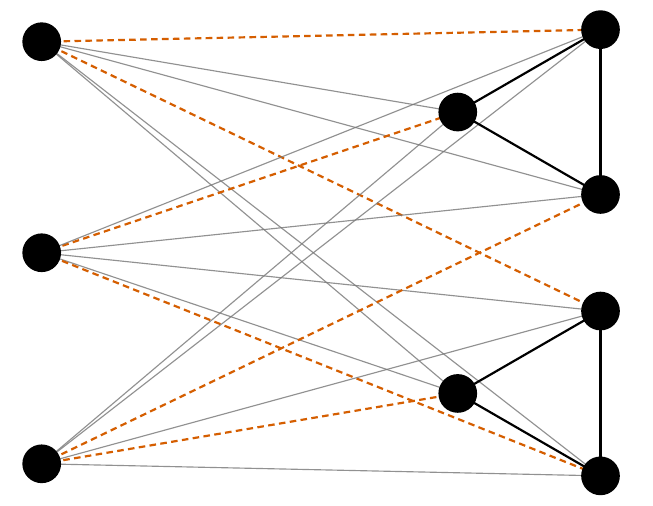}
    \caption{A schematic illustration of the CK graph ($G_m,m=3$) and its modifications ($G'_m,m=3$) of the MIS problem. Solid black circles represent vertices, and lines represent edges. In the CK graph $G_3$, the left (isolated dots) and right (triangles) sections are fully connected (both solid and dashed lines). Its MIS consists of the three vertices on the left, while the triangles on the right generate competing local minima. A CK-like graph $G'_3$ is obtained by randomly removing connections (orange dashed lines) between the $G_3$'s two sides. }
    \label{fig:CK_CK_like_schematic}
\end{figure}

Given a graph $G=(\mathcal V, \mathcal E)$, the maximum independent set (MIS) problem consists of $n=|\mathcal{V}|$ vertices $\mathcal{V}$ and a series of connected edges $\mathcal E$. The goal is to find the largest subset of $\mathcal{V}$ whose pairwise edges are not in $\mathcal E$ \cite{Karp1972, GareyJohnson1979, TarjanTrojanowski1977}. In other words, the goal is to find the binary configuration $x\in\{0,1\}^{|\mathcal V|}$ that maximizes $\sum_i x_i$, where $x_i=1$ ($x_i=0$) indicates that vertex $i$ is selected (unselected), while satisfying the constraint $x_i x_j=0$ for every $(i, j)\in \mathcal E$. In particular, we consider the CK graph family $G_m$ and its variant, the CK-like graph family $G'_m$, for numerical demonstration. The graph $G_m$ consists of $m$ disconnected vertices on the left and $m-1$ disconnected triangles on the right, mutually fully connected, totaling $n=4m-3$ vertices and $|\mathcal E|=3(m^2-1)$ edges. For the original CK graph family $G_m$, the MIS solution is all $x_i=1$ ($x_i=0$) on the left (right), whereas numerous sub-optimal independent sets that cling to the triangles on the right create difficult barriers to optimization. Consequently, the conventional transverse-field QAA exhibits an exponentially closing gap minimum in this graph family \cite{AminChoi2009, Choi2010}, making it a hard benchmark for testing the effectiveness and robustness as we extend QAA into the non-Hermitian realm. $G'_m$ is obtained from $G_m$ by randomly removing part of the edges between the left vertices and the triangles on the right. $G_3$ and $G'_3$ are illustrated in Fig.~\ref{fig:CK_CK_like_schematic}.

\begin{figure}[t]
    \centering
    \includegraphics[width=0.95\linewidth]{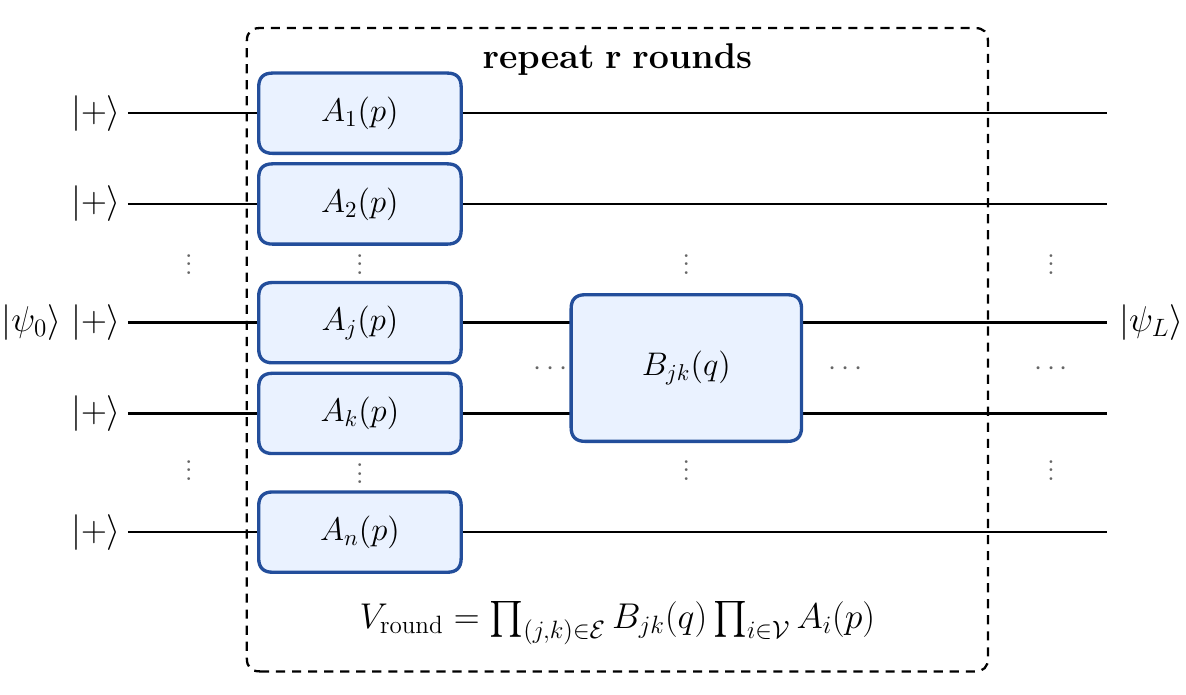}
    \caption{A schematic illustration of the non-unitary QC for solving the MIS problem: given an initial state $|\psi_0\rangle$ with equal-weight superposition $|+\rangle^{\otimes n}$, the output is obtained after repeated $r$ rounds of operations. Within each round (dashed box), vertex-amplification gates $A_i(p)$ and two-qubit edge-constraint gates $B_{jk}(q)$ are applied according to the vertices and edges of the MIS graph. }
    \label{fig:MIS-NHQC_schematic}
\end{figure}

First, we propose a non-unitary QC using simple, diagonal non-unitary gates to solve a target MIS problem. We introduce two elementary gates: a vertex-amplification gate $A_i(p)$ and a two-qubit edge-constraint gate
$B_{jk}(q)$: 
\begin{equation}
A_i(p) = c_A
\begin{pmatrix}
1 & 0\\
0 & p
\end{pmatrix}_i,
\quad
B_{jk}(q) = c_B
\begin{pmatrix}
q & 0 & 0 & 0\\
0 & q & 0 & 0\\
0 & 0 & q & 0\\
0 & 0 & 0 & 1
\end{pmatrix}_{jk},
\label{eq:ABmatrix}
\end{equation}
which relatively amplifies $|1\rangle_i$, the selection of the vertex $i$ by a factor $p$, and $|00\rangle_{jk}$, $|01\rangle_{jk}$, and $|10\rangle_{jk}$, the configurations satisfying the independent-set constraint on edge $jk$ by a factor $q$, respectively, with the requirement that $q>p>1$. $c_A$ and $c_B$ are positive scalars for overall normalization, which, without loss of generality, we set as $c_A=c_B=1$ unless stated otherwise. Clearly, $[A_i, B_{jk}]=0$ for arbitrary $i$, $j$, and $k$. Our QC consists of $r$ rounds of $V_{\mathrm{round}}=\prod_{(j, k)\in \mathcal{E}}B_{jk}(q)\prod_{i\in\mathcal{V}}A_i(p)$, totaling $L = r (|\mathcal{E}|+|\mathcal{V}|)$ steps, and the initial state $|\psi_0\rangle = |+\rangle^{\otimes n}$ is an equal-weight superposition of all candidate vertex configurations. Schematically, the non-unitary QC is illustrated in Fig.~\ref{fig:MIS-NHQC_schematic}. 

This non-unitary QC relatively reweighs each configuration $x$ with a factor $p^{2rF(x)}$, where: 
\begin{equation}
F(x)=|x|-\lambda \sum_{(i,j)\in \mathcal{E}}x_i x_j, \quad \lambda=\frac{\log q}{\log p}>1. 
\end{equation}
Clearly, the MIS has the largest $F(x)$. In particular, for the CK graph family $G_m$, the number of rounds to reach a fixed success probability for the target MIS scales linearly with $m$, $r=\Theta(m)$. Therefore, the total number of gates scales cubically with the problem size $n$: 
\begin{equation}
L=r(|\mathcal{V}|+|\mathcal{E}|)=\Theta(m^3).
\end{equation}
Further algorithmic details of and analysis on the non-unitary QC for the CK graph family are given in Appendix. 

Next, we convert the above QC into an HD QAA. Following Sec. \ref{sec:history_decoupled_mapping}, the resulting Hamiltonian during the $l$-th interval, $l = 1, 2, \cdots, L$, takes the following form in the composite space: 
\begin{equation}
H_l(s)
=
\frac{\Omega}{2}
\begin{pmatrix}
    1-\cos 2\theta(s)
    &
    -\sin 2\theta(s)\,V_l^{-1}
    \\
    -\sin 2\theta(s)\,V_l
    &
    1+\cos 2\theta(s)
\end{pmatrix},
\label{eq:HD_block}
\end{equation}
where the first and second row (column) denote the $|l-1\rangle$ and $|l\rangle$ states in the auxiliary clock space $\mathcal{H}_c$, and depending on $l$, $V_l=A_i(p)$ acts on a single qubit or $V_l=B_{jk}(q)$ acts on two qubits in the work space $\mathcal{H}_w$, see Eq. \eqref{eq:ABmatrix}. Equivalently, the Hamiltonian can be written as:
\begin{eqnarray}
    H_l(s) &=& \Omega \left[\frac{I}{2} - \frac{\cos 2\theta(s)}{2}\tau^{(l)}_z - \frac{(p+1)^2\sin 2\theta(s)}{8p}\tau^{(l)}_x \right. \nonumber\\
    & &  + \frac{(p-1)^2\sin 2\theta(s)}{8p}\tau^{(l)}_x\sigma^z_i \nonumber\\
    & &  + \left. \frac{i(p^2-1)\sin 2\theta(s)}{8p}\tau^{(l)}_y(1-\sigma_i^z) \right],  
\end{eqnarray}
if $V_l=A_i(p)$, and: 
\begin{eqnarray}
    H_l(s) &=& \Omega \left[\frac{I}{2} - \frac{\cos 2\theta(s)}{2}\tau^{(l)}_z - \frac{(q+1)^2\sin 2\theta(s)}{8q}\tau^{(l)}_x \right. \nonumber\\
    & &  + \frac{(q-1)^2\sin 2\theta(s)}{16q}\tau^{(l)}_x (\sigma^z_j\sigma^z_k-\sigma^z_j-\sigma^z_k-1)\nonumber\\
    & &  + \left. \frac{i(q^2-1)\sin 2\theta(s)}{16q}\tau^{(l)}_y(3+\sigma_j^z+\sigma_k^z-\sigma_j^z\sigma_k^z) \right], \nonumber\\ 
\end{eqnarray}
if $V_l=B_{jk}(q)$, where $\tau^{(l)}_x$, $\tau^{(l)}_y$, and $\tau^{(l)}_z$ are Pauli matrices acting on the $|l-1\rangle$ and $|l\rangle$ states in the clock space, and $\sigma^z_i$ act on the $i$-th qubit in the work space. In either case, the last term in the Hamiltonian is manifestly non-Hermitian. 

Ideally, such a Hamiltonian path should evolve an initial ground state $|\psi_0\rangle=|+\rangle^{\otimes n}\otimes |0\rangle$ to the final ground state adiabatically: 
\begin{equation}
|\psi_L\rangle \propto (V_{\mathrm{round}})^r |+\rangle^{\otimes n} \otimes |L\rangle, 
\label{eq:reffect}
\end{equation}
which harbors the MIS solution as a configuration with sufficient weight to be measured. 

\subsection{Numerical Results} 
\label{subsec:numerical_experiment}

To showcase the effectiveness and stability of the HD QAA, we summarize QAA results on MIS problems, including $G_m$ and $G'_m$ as discussed in the previous subsection and Fig. \ref{fig:CK_CK_like_schematic}. In particular, given a specific MIS problem, we derive its corresponding non-unitary QC and map it to a non-Hermitian Hamiltonian path, as we discussed in Subsec.~\ref{subsec:nh_mis_algorithm}. Hereafter, without loss of generality, we set $p=2$, $q=4$, and $\Omega=1$, unless noted otherwise. For comparison, we consider, in parallel, the FK QAA, as well as a Hermitian QAA (HM QAA) following the Grover algorithm; see details of the HM QAA in the Appendix. Initiating with a trivial ground state $|\psi(0)\rangle=|+\rangle^{\otimes n}\otimes |0\rangle$, we keep track of the spectral gaps and the success probability, the weight of the MIS configuration $|x_{MIS}\rangle$ in the clock-state $|L\rangle$ sector of the final state: 
\begin{equation}
P_{MIS} = \frac{\left\langle
\Psi(T)\right| 
\left(|x_{MIS} \rangle\langle x_{MIS}| \otimes |L\rangle\langle L| \right)
\left| \Psi(T) \right\rangle }{\left\langle
\Psi(T)|\Psi(T) \right\rangle}. 
\label{eq:ideal_suss}
\end{equation}

\begin{figure}[t]
    \centering
    \includegraphics[width=\columnwidth]{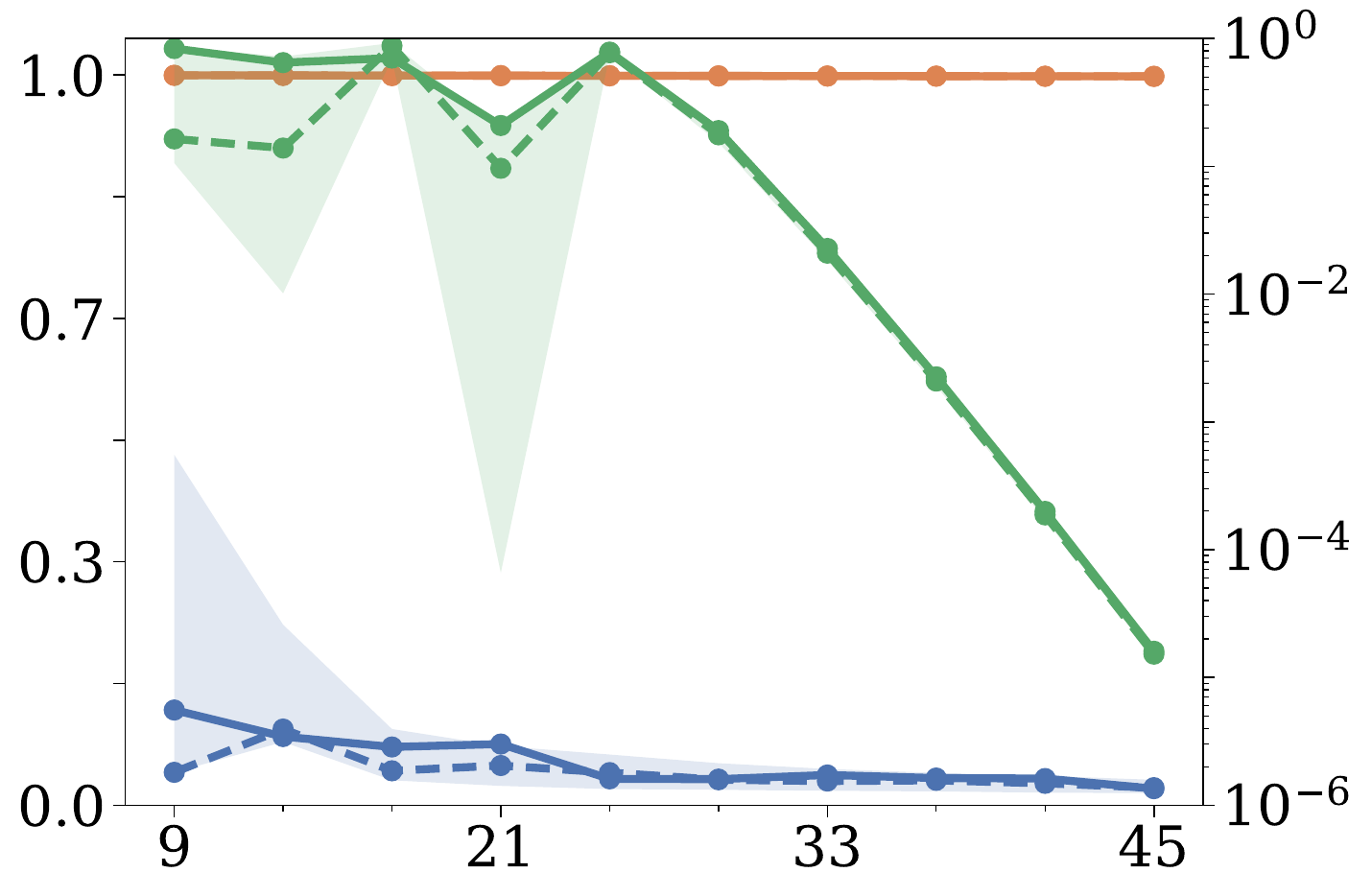}
    \caption{The success probability $P_{MIS}^{(0)}$ of the three QAAs exhibits qualitatively different scalings versus the problem size $n$. The HM QAA (green, right log axis) suffers an exponential decrease, whereas the non-Hermitian QAAs' $P_{MIS}^{(0)}$ (left linear axis) decrease much more slowly. Note the difference in the vertical axis. Still, the HD QAA (orange) leads with a success probability $P_{MIS}^{(0)}\sim 1$ versus the polynomial scaling of the FK QAA (blue). Here, the solid lines indicate results on the CK graphs $G_m$, while the dashed lines and shaded regions denote the medians and the ranges across 200 CK-like graphs $G'_m$. }
    \label{fig:experiment1_scaling_gamma10}
\end{figure}

First, we study the relationship between the success probability $P_{MIS}^{(0)}$ and the size of the MIS problem $n$ for a finite evolution time $T$ without any noise or perturbation. For each QAA,  we solve the corresponding time-dependent Schr\"odinger equation: 
\begin{equation}
\label{eq:clean_time_dependent_sch_eq}
i\frac{d}{dt}|\Psi(t)\rangle = H(t) |\Psi(t)\rangle. 
\end{equation}
For fair comparison, we fix the total evolution time $T=10L$ to be identical across the three QAAs; also, for consistent success probability, we adopt $r=n$; and thus, $T=\frac{5n}{8}(3n^2+34n-21)$ for both non-Hermitian QAAs and the same number of Grover iterations and time for the HM QAA. 

We summarize the resulting success probability $P_{MIS}^{(0)}$ over a range of problem size $n=9, 13, 17,\ldots, 45$ in Fig.~\ref{fig:experiment1_scaling_gamma10}. The superscript $(0)$ denotes the absence of perturbation at this point. Clearly, HD QAA maintains a high success probability $\sim1$; FK QAA also achieves a notable $P_{MIS}^{(0)}$, which may suffer a moderate polynomial decay as $n$ increases. In contrast, the HM QAA witnesses an exponentially declining $P_{MIS}^{(0)}$ as $n$ increases, consistent with the obstacle that a Hermitian Grover-type algorithm cannot significantly amplify the MIS within polynomial depth.

\begin{figure}[t]
    \centering
    \includegraphics[width=\columnwidth]{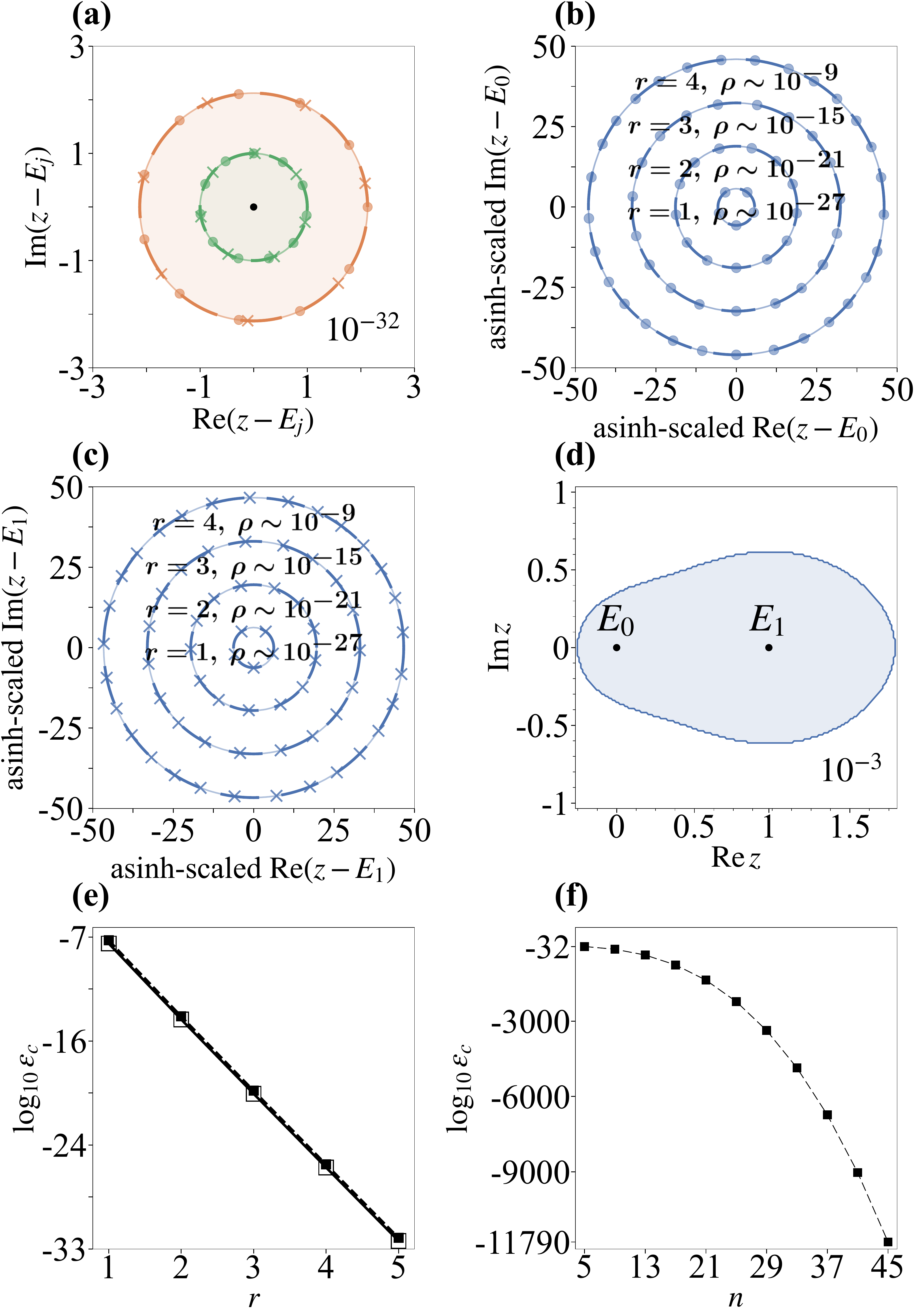}
\caption{Pseudospectral and gap stability of the FK QAA (blue), HD QAA (orange), and HM QAA (green) for the CK-graph $G_m$ of the MIS problem against perturbations $\varepsilon$ for various numbers $r$ of rounds of operations. (a) While the pseudospectra remain negligible for both HD and HM QAAs, regardless of $E_0$ or $E_1$ states, $r = 1,\ldots,5$, its contours for the FK QAA around (b) $E_0$ and (c) $E_1$ expand exponentially as $r=1, 2, 3, 4$ increases, and (d) merge between $E_0$ and $E_1$ at $r=5$, indicating gap closure. Here,  $\varepsilon=10^{-32}$, the dashed lines are theoretical estimates using Eq. \eqref{eq:rhoestimate}, and the solid lines with hollow circles (crosses) represent the numerical results for $E_0$ ($E_1$). For clarity, we have plotted the radial scale in (b) and (c) asymptotically logarithmic as $\operatorname{asinh}(\rho/10^{-29})$, $\rho = |z-E_j|$. (e) The threshold perturbation strength $\varepsilon_c$, where the gap of the FK QAA closes, decreases exponentially as the operation round $r$. The dashed (solid) line with solid (hollow) squares represents the theoretical estimates (numerical results). We have set $m=2$ for (a)-(e). (f) Theoretical estimate of the threshold perturbation strength $\varepsilon_c$ for FK QAA on larger CK graphs $G_m$, where $r=n=4m-3$. }
\label{fig.exp_two_psedospectrum}
\end{figure}

On the other hand, the FK QAA is not pseudospectrally stable. More directly, to diagnose this instability, we evaluate the pseudospectra around the ground state $E_0$ and the first excited state $E_1$ for representative Hamiltonians of the three QAAs along their evolution paths; see Appendix for further details of the setup. Without loss of generality, we summarize the results on the CK graph with $n=5$ ($m=2$) through high-precision numerical algorithms in Fig.~\ref{fig.exp_two_psedospectrum}. 

Under a fixed, small perturbation strength $\varepsilon=10^{-32}$, we derive the pseudospectral regions around $E_0$ and $E_1$ following Eq.~\eqref{eq:resolvent_pesedospectrum} as the number of algorithmic rounds $r=1, 2, 3, 4, 5$ increases, which is essential for larger problems and better outcomes - Eq. \eqref{eq:reffect}. Clearly, the pseudospectra remain stable for the HM QAA and the HD QAA Figs.~\ref{fig.exp_two_psedospectrum}(a); however, they expand rapidly and eventually merge $E_0$ and $E_1$ at $r=5$ for the FK QAA [Figs.~\ref{fig.exp_two_psedospectrum}(b),(c) and (d)], indicating gap closure and failure of the adiabatic condition even under such small perturbations $\varepsilon$ and short process $r$. More quantitatively, in Fig.~\ref{fig.exp_two_psedospectrum}(e), we show the threshold of such perturbation $\varepsilon_c$ that merges the pseudospectral regions around $E_0$ and $E_1$, which exhibits an exponential decrease versus $r$. Also, these results are fully consistent with the theoretical estimates according to Eq.~\eqref{eq:rhoestimate}. 

Finally, we estimate the perturbation threshold $\varepsilon_c$ for larger CK graphs $G_m$ with $m=2, \ldots, 7$ and the FK QAA with $r=n=4m-3$: 
\begin{equation}
\label{eq:gap_close_}
	\varepsilon_c  \simeq \frac{E_1-E_0}{\|\Pi_0\|+\|\Pi_1\|}, 
\end{equation}
where the pseudospectra around $E_0$ and $E_1$ begin to merge. We calculate $E_0$, $E_1$ and their projectors $\Pi_0$, $\Pi_1$ for $H(s)$ associated with the MIS solution, $x=x_{\rm MIS}$, where the cumulative amplification induced by $W_L$, which is proportional to $p^{rF(x)}$, is maximal and the pseudospectral problem is the most prominent. The results in Fig.~\ref{fig.exp_two_psedospectrum}(f) show that $\varepsilon_c$ decreases exponentially with $n$, indicating an exponentially shrinking tolerance to perturbations and increasingly severe pseudospectral instability; see further details in Appendix. Therefore, proper pseudospectral stability is paramount for scalable non-Hermitian QAA. 

\begin{figure}[t]
    \centering
    \includegraphics[width=\columnwidth]{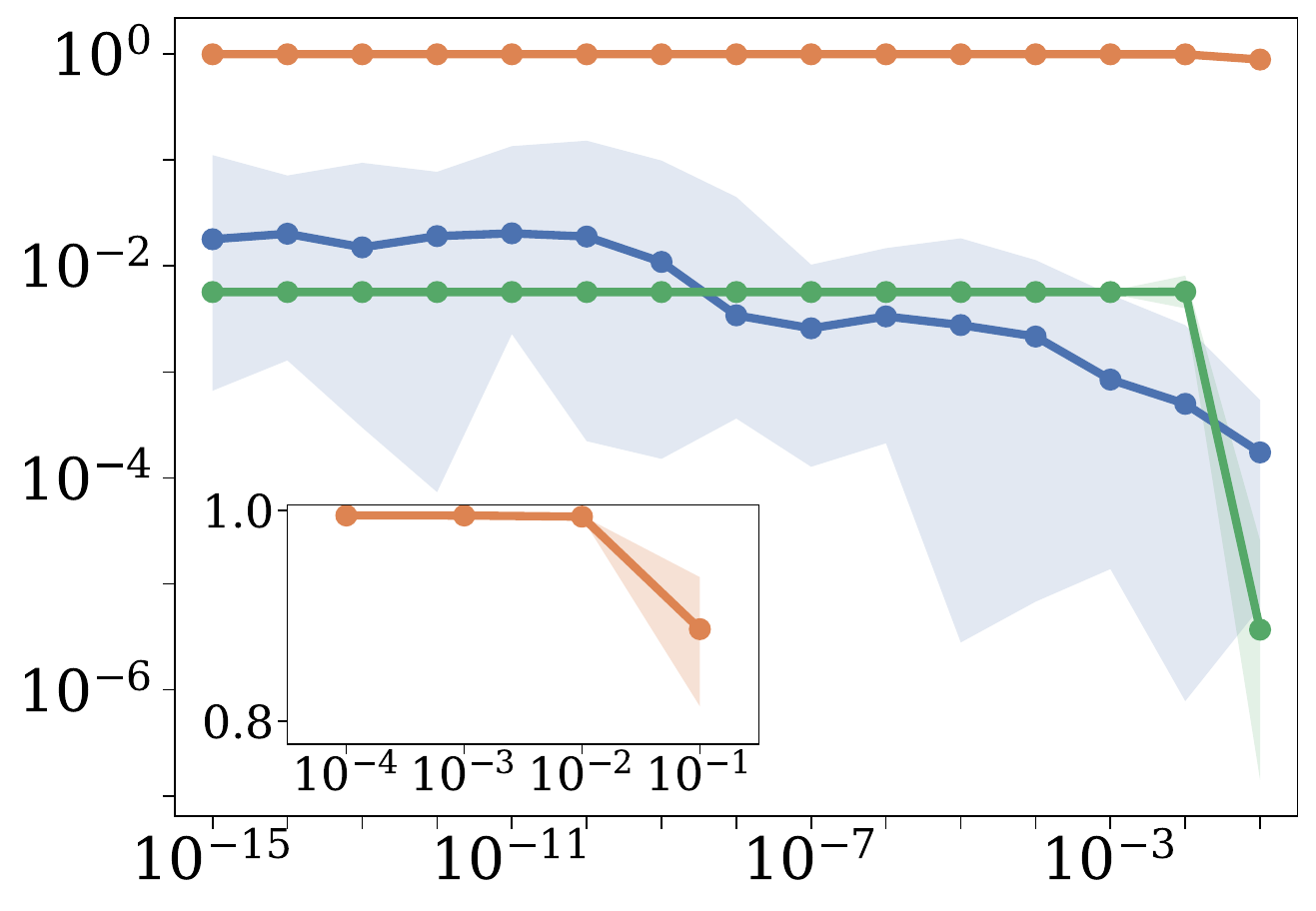}
    \caption{The success probability $P_{MIS}^{(\varepsilon)}$ of the three QAAs for the CK graph $G_2$ of the MIS problem versus perturbation strength $\varepsilon$: the HD QAA (orange) exhibits outstanding robustness, while the FK QAA (blue) and HM QAA (green) both witness notable algorithmic failure. The inset is an enlargement over large $\varepsilon$. The shaded regions indicate distributions over 100 independent random noises, with their time dependence represented by 70 independent instances spanning the entire evolution. }
    \label{fig:mis_noisy_dynamics} 
    \end{figure}

Indeed, as we include the perturbation $\Delta H(t) = \varepsilon \frac{R}{\|R\|_2}, $ into the time-dependent Schr\"odinger equation in Eq. \ref{eq:clean_time_dependent_sch_eq} [$R$ is a complex random matrix with the same dimension as the original QAA Hamiltonian, so that $\|\Delta H(t)\|_2=\varepsilon$]: 
\begin{equation}
i\frac{d}{dt}|\Psi(t)\rangle = (H(t)+\Delta H(t)) |\Psi(t)\rangle, 
\label{eq:time_dependent_sch_eq}
\end{equation}
the $P_{MIS}^{(\varepsilon)}$ for the CK graph $G_2$ of the MIS problem is summarized in Fig.~\ref{fig:mis_noisy_dynamics}. The rest of the settings are identical to Fig. \ref{fig:experiment1_scaling_gamma10}.

Clearly, over the entire test range, the success probability $P_{MIS}^{(\varepsilon)}$ for the HD QAA does not show much deterioration or distinction from the pristine case $P_{MIS}^{(0)}$ without any noise or perturbation. Likewise, the HM QAA remains insensitive to small perturbations and only exhibits a noticeable decrease in $P_{MIS}^{(\varepsilon)}$ at sufficiently large perturbations $\varepsilon=10^{-1}$. On the other hand, $P_{MIS}^{(\varepsilon)}$ for the FK QAA decreases significantly and exhibits poor stability as the noise strength increases. Therefore, across the board, only the HD QAA has demonstrated satisfactory effectiveness and robustness in solving typical and difficult MIS problems, as summarized in Table~\ref{tab:qaa_comparison}.

\subsection{Potential realizations} \label{subsec:Potential_realizations}

In this subsection, we discuss a potential realization of the above non-Hermitian HD QAA scheme based on coupled optical waveguides. Because our choices of $V_l \in \{A_i(p), B_{jk}(q)\}$ are fully diagonal in configuration basis $x$: 
\begin{equation}
A_i(p)|x\rangle=c_Ap^{x_i}|x\rangle,
\qquad
B_{jk}(q)|x\rangle=c_Bq^{1-x_jx_k}|x\rangle,
\end{equation}
where we set $c_A=p^{-1/2}$ and $c_B=q^{-1/2}$ for the optical realization. We can decouple the Hamiltonian in Eq.~\eqref{eq:HD_block} into $2\times2$ blocks in the $\{|l-1\rangle, |l\rangle\}$ clock space for each independent $x$: 
\begin{align}
H_l(s)
&=
\bigoplus_x H_{x,l}(s),
\\
H_{x,l}(s)
&=
\frac{\Omega}{2}
\begin{pmatrix}
    1-\cos 2\theta(s)
    &
    -\sin 2\theta(s)\,v_l(x)^{-1}
    \\
    -\sin 2\theta(s)\,v_l(x)
    &
    1+\cos 2\theta(s)
\end{pmatrix},
\label{eq:Hlx}
\end{align}
where $v_l(x)=p^{x_i-1/2}$ for $V_l=A_i(p)$ and $v_l(x)=q^{1/2-x_jx_k}$ for $V_l=B_{jk}(q)$. Interestingly, such a non-Hermitian $H_{l, x}(s)$ can be realized through (a series of) optical systems of three neighboring waveguides with loss and couplings.

\begin{figure}
    \centering
    \includegraphics[width=1\linewidth]{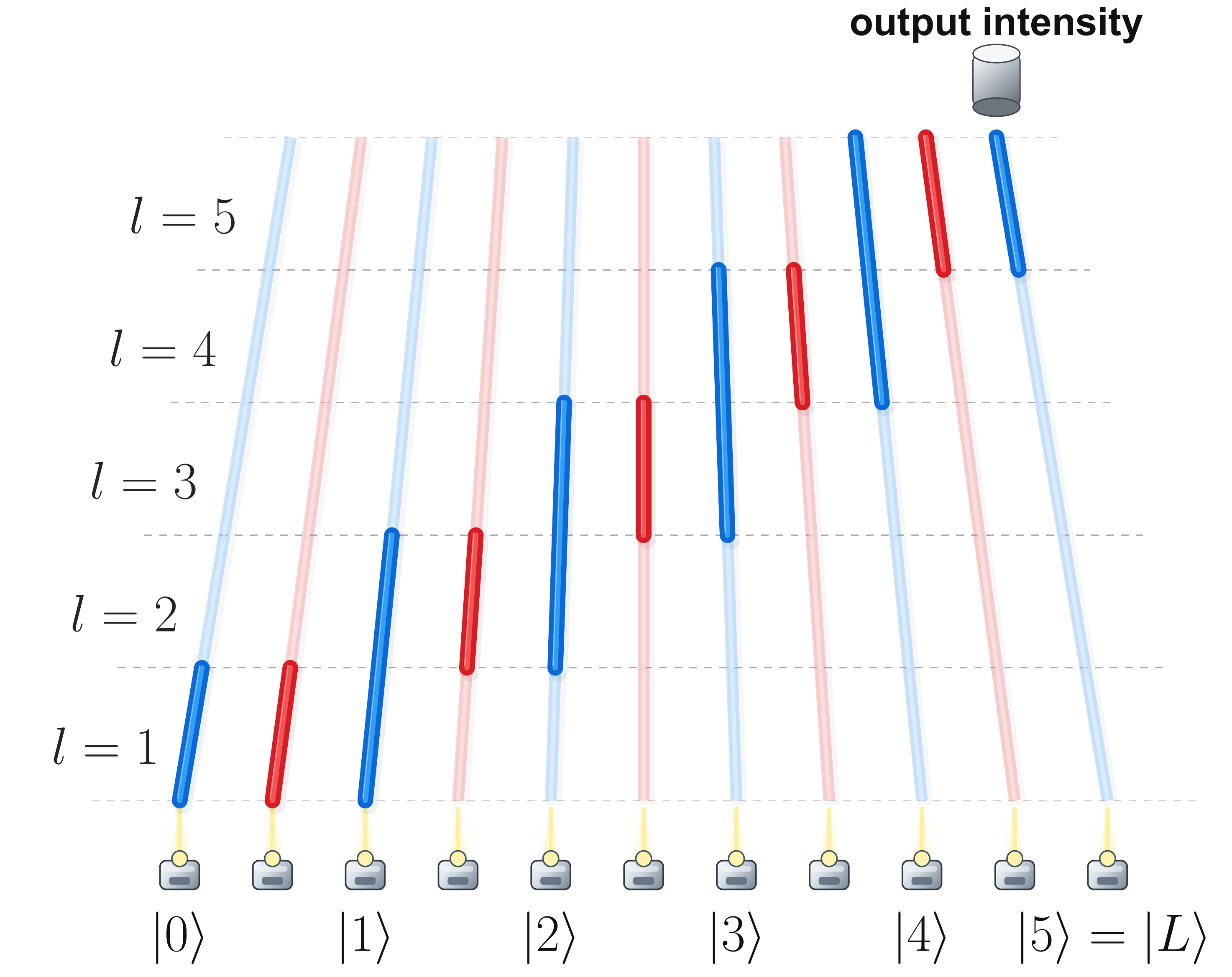}
    \caption{A schematic illustration of the proposed coupled-waveguide implementation of the HD QAA: the blue waveguides correspond to the decoupled, fixed configuration $x$ in the workspace and states $|0\rangle$, $|1\rangle$, $\ldots$, $|L\rangle$, respectively, in the clock space; the red waveguides are dissipative auxiliary units to introduce non-Hermiticity. Light propagates along the $z$ direction, simulating time-dependent dynamics, and is further divided into $L$ intervals: only couplings between waveguides $|l-1\rangle$ and $|l\rangle$, along with the associated auxiliary waveguide, are activated according to the HD QAA Hamiltonian in the $l$-th interval. The optical amplitude is gradually transferred from the first waveguide to the second, and so on, until the intensity at the last waveguide is measured, yielding the output for configuration $x$. The MIS corresponds to the maximum output. Here, $L=5$.} 
    \label{fig:Schematic_Realizations}
\end{figure}

Specifically, for a configuration $x$, we consider the coupled-mode equation \cite{Ruter2010PT, Regensburger2012PTLattices, MiriAlu2019} along a propagation direction $z$: 
\begin{equation}
\resizebox{1\columnwidth}{!}{$\displaystyle
i\frac{d}{dz}
\begin{pmatrix}
\mu_{x,l}\\
\nu_{x,l}\\
\xi_{x,l}
\end{pmatrix}
=
\begin{pmatrix}
\beta_{x,l}^{\mu}(z) & J_{x,l}(z) & g_{x,l}^{\mu}(z)\\
J_{x,l}(z) & \beta_{x,l}^{\nu}(z) & g_{x,l}^{\nu}(z)\\
\left[g_{x,l}^{\mu}(z)\right]^*
&
\left[g_{x,l}^{\nu}(z)\right]^*
&
\beta_{x,l}^{\xi}(z)-i\gamma
\end{pmatrix}
\begin{pmatrix}
\mu_{x,l}\\
\nu_{x,l}\\
\xi_{x,l}
\end{pmatrix},
$}
\label{eq:optical}
\end{equation}
where the optical fields in the three waveguides:  
\begin{equation}
\Psi_{x,l}(z)= \begin{pmatrix}
\mu_{x,l}\\ \nu_{x,l}\\ \xi_{x,l} 
\end{pmatrix}, 
\end{equation}
represent the two clock states, $|l-1\rangle$ and $|l\rangle$, and an auxiliary lossy waveguide $\xi$ for introducing non-Hermiticity. The diagonal terms $\beta_{x,l}^{\mu}(z)$, $\beta_{x,l}^{\nu}(z)$, and $\beta_{x,l}^{\xi}(z)$ are the propagation constants (optical fields' phase accumulation rates) of the three respective waveguides: 
\begin{equation}
\beta_{x,l}^{\alpha}(z)=k_0 n_{\mathrm{eff},x,l}^{\alpha}(z), \quad k_0=2\pi/\lambda_0,
\quad
\alpha\in\{\mu,\nu,\xi\} 
\end{equation}
where $\lambda_0$ ($k_0=2\pi/\lambda_0$) is the operating vacuum wavelength (wavenumber), and $n_{\mathrm{eff}, x,l}^{\alpha}(z)$ is the effective refractive index of the waveguide labeled by $\alpha$. We choose a common propagation constant
$\beta_{x,l}^{\xi}(z)=\beta_0$ for all auxiliary waveguides and work
in the corresponding rotating frame, which changes only an overall
phase and does not affect the output intensity. The
propagation-constant detunings are then: 
\begin{equation}
\chi_{x,l}^{\alpha'}(z)
\equiv
\beta_{x,l}^{\alpha'}(z)-\beta_0,\qquad
\alpha'\in\{\mu,\nu\}.
\end{equation}
The subscript $x$ labels the configuration in the workspace $\mathcal {H} _ {w} $, while $l$ keeps track of not only the QAA process but also the active waveguides, corresponding to the $|l-1\rangle$ and $|l\rangle$ clock states; see Fig. \ref{fig:Schematic_Realizations}. The off-diagonal terms, $J_{x,l}(z)$, $g_{x,l}^{\mu}(z)$, and $g_{x,l}^{\nu}(z)$, describe reciprocal couplings between the waveguides, commonly arising from the overlap of their evanescent fields. We may tune their magnitudes and phases by varying waveguide separations
and refractive indices or through synthetic gauge-field engineering
\cite{Chen2021WaveguideTB, Fang2012GaugeField}. Finally, the $-i\gamma$ term represents a loss in the lossy auxiliary waveguide, where $\gamma>0$ is the loss rate. Hereafter, we will suppress the $x$ labels for simplicity.

Eq. \eqref{eq:optical} has a similar form to the Schrödinger equation, except that the evolution is not in time $t$ but instead in the coordinate $z$ in the propagation direction. Thus, we can identify the HD QAA interval $l$ and normalized time $s\in[0,1]$ as: 
\begin{equation}
s=\frac{z-z_{l-1}}{Z_l}, \quad z\in[z_{l-1},z_{l-1}+Z_l], 
\end{equation}
where $Z_l$ is the propagation length corresponding to the $l$-th interval, $z_l\equiv z_{l-1}+Z_l$ denotes its endpoint (with $z_0=0$), and the total HD-QAA evolution time $T$ is implemented by the total propagation length $Z_{\mathrm{tot}}=\sum_{l=1}^{L}Z_l$. In short, with proper parameterization, the propagation of light through this waveguide system may simulate the adiabatic evolution along the entire HD-QAA path. 

In particular, in the strong-loss limit,
\begin{equation}
\label{eq:large_gamma}
\gamma \gg
\max\left\{
|g_l^\mu(z)|,\,
|g_l^\nu(z)|,\,
|J_l(z)|,\,
|\chi_l^\mu(z)|,\,
|\chi_l^\nu(z)|
\right\},
\end{equation}
together with the assumed slow modulation along $z$, the optical field in the auxiliary waveguide does not accumulate over long distances, but rapidly reaches a steady state. Therefore, we have approximately:
\begin{equation}
\label{eq:gamma_gg}
\frac{d\xi_{l}}{dz}\simeq 0, \quad \xi_{l}(z)\simeq -\frac{i}{\gamma} \left[ \left[g_{l}^{\mu}(z)\right]^*\mu_{l}(z)+\left[g_{l}^{\nu}(z)\right]^*\nu_{l}(z) \right], 
\end{equation}
to the leading order. Consequently, the effective equation across the two main waveguides is written as:
\begin{equation}
i\frac{d}{dz}
\begin{pmatrix}
\mu_{l}\\ \nu_{l}
\end{pmatrix} 
= H_{\mathrm{eff},l}(z)
\begin{pmatrix}
\mu_{l}\\ \nu_{l}
\end{pmatrix},
\end{equation}
where $H_{\mathrm{eff},l}(z)$ is given by:
\begin{equation}
\begin{pmatrix} 
\chi_{l}^{\mu}(z)-i|g_{l}^{\mu}(z)|^2/\gamma & J_{l}(z)-i g_{l}^{\mu}(z)\left[g_{l}^{\nu}(z)\right]^*/\gamma \\
J_{l}(z)-i g_{l}^{\nu}(z)\left[g_{l}^{\mu}(z)\right]^*/\gamma & \chi_{l}^{\nu}(z)-i|g_{l}^{\nu}(z)|^2/\gamma
\end{pmatrix},
\end{equation}
which replicates \eqref{eq:Hlx} with proper parameterizations as follows. First, we set the two couplings to an equal amplitude: 
\begin{equation}
|g_{l}^{\mu}(z)|=|g_{l}^{\nu}(z)|=g_{l}(z),
\end{equation}
with a relative phase difference: 
\begin{eqnarray}
g_{l}^{\mu}(z)&=&g_{l}(z)e^{i\phi_{l}^{\mu}(z)}, \nonumber \\
g_{l}^{\nu}(z)&=&g_{l}(z)e^{i\phi_{l}^{\nu}(z)}, \nonumber \\
\phi_{l}^{\nu}(z)-\phi_{l}^{\mu}(z)&=&
\begin{cases}
-\pi/2, & v_l(x)>1,\\
+\pi/2, & v_l(x)<1.
\end{cases}
\label{eq:phase_matching}
\end{eqnarray}
Besides, we set the third coupling between the two main waveguides as: 
\begin{equation}
J_{l}(z)
=
-\frac{\Omega}{4}
\sin 2\theta(z)
\left[
v_l(x)+v_l(x)^{-1}
\right],
\label{eq:J_matching}
\end{equation}
and the detunings as: 
\begin{equation}
\chi_{l}^{\mu}(z)
=
\frac{\Omega}{2}
\left[1-\cos 2\theta(z)\right],
\quad
\chi_{l}^{\nu}(z)
=
\frac{\Omega}{2}
\left[1+\cos 2\theta(z)\right],
\label{eq:detuings_matching}
\end{equation}
and, finally, a loss rate $\gamma$ that obeys:
\begin{equation}
\frac{g_{l}^2(z)}{\gamma}
=
\frac{\Omega}{4}
\sin 2\theta(z)
\left|
v_l(x)-v_l(x)^{-1}
\right|.
\label{eq:g_matching}
\end{equation}
Consequently, $H_{\mathrm{eff},x,l}(z)$ reproduces $H_{x,l}(s)$ up to a common scalar attenuation.

Globally, we introduce $L+1$ waveguides (with $L$ other auxiliary waveguides) arranged in the transverse direction, representing the clock states $|0\rangle$, $|1\rangle$, \ldots, $|L\rangle$, while the optical field propagates along their common longitudinal $z$ direction; see illustration in Fig.~\ref{fig:Schematic_Realizations}. As the state propagates and evolves along $z$ and reaches $z_l$, i.e., the end of a QAA's $l^{th}$ interval, it gradually and adiabatically transfers its weight from across the neighboring $(l-1)^{th}$ and $l^{th}$ waveguides to solely on the $l^{th}$ waveguide, according to the previous paragraphs. As the next [$(l+1)^{th}$] interval starts, the active couplings switch gear and move on between the neighboring $l^{th}$ and $(l+1)^{th}$ waveguides, and so on, in a continuous and matching fashion: $\nu_{l}\equiv\mu_{l+1}$. For the other waveguides not actively involved in the current interval, $l'\neq l-1, l$, we can set $\chi_{l'}^{\alpha}(z)=\Omega$ without couplings $g_{l'}^{\alpha}(z)=J_{l'}(z)=0$, for a trivial propagation $H=\Omega I_{\mathrm{inact}}$. 

Intuitively, when the optical field propagates along the $z$ direction, its amplitude is gradually transferred from the $|0\rangle$ waveguide to the $|1\rangle$ waveguide, it acquires a non-unitary factor $v_1(x)$; subsequently, when it is transferred from the $|1\rangle$ waveguide to the $|2\rangle$ waveguide during the next interval, it acquires a factor $v_2(x)$; proceeding interval by interval, $|0\rangle \to |1\rangle \to \cdots\to |L\rangle$,
the output amplitude from the last waveguide accumulates an overall factor $\nu_{L}\propto v_L(x) \cdots  v_2(x)v_1(x)\mu_{1}(z_0)$. Therefore, such a waveguide system can simulate the adiabatic evolution associated with a single configuration $x$. In practice, the output intensities $|\nu_{L}|^2$ can be measured in parallel for all $x$ configurations, and the MIS solution corresponds to the largest output intensity. For an $n$-qubit problem, the complete HD QAA can, in principle, be implemented by $2^{n}$ parallel systems, totaling $2^{n}(2L+1)$ waveguides.

\section{Conclusion and Discussion} \label{sec:conclusion}

Non-Hermitian QAAs offer an alternative quantum computation paradigm with potential advantages in robustness, scalability, and physical overhead. In this work, we have first clarified that a practical non-Hermitian QAA requires not only a finite spectral gap but also a real spectrum and a stable pseudospectrum. For finite-dimensional diagonalizable non-Hermitian Hamiltonians, these conditions amount to the existence and singular features of an invertible similarity transformation from the QAA Hamiltonian to a Hermitian Hamiltonian. In particular, based on the equivalent definitions of the pseudospectrum in Eqs.~\eqref{eq:resolvent_pesedospectrum} and \eqref{eq:perturbations_pseudospectrum}, we characterize the QAA's stability against perturbations using the resolvent expansion in Eq.~\eqref{eq:rhoestimate} and the condition number in Eq.~\eqref{eq:condition_number_estimate}, respectively. Our analytical and numerical results indeed demonstrate that a real and gapped spectrum alone is insufficient to guarantee adiabatic evolution. 

Then, to obtain such a non-Hermitian QAA - a continuous-time Hamiltonian evolution, we take advantage of the straightforward and efficient optimization of non-unitary QC. However, through a direct generalization of the Hermitian QC-to-QAA mapping, i.e., the FK construction, the resulting QAA's real and gapped spectrum is not pseudospectrally stable. Therefore, we have introduced the HD QAA in
Eq.~\eqref{eq:HD_Hamiltonian}, which decouples the instantaneous Hamiltonian from the past non-unitary gates, thereby guaranteeing both a real, gapped spectrum and a controlled pseudospectrum. Its final state, which is entirely in the $|L\rangle$ sector, and its constant gap also make it more efficient than the FK QAA in theory. 

For demonstration, we consider the non-Hermitian QAA for the MIS problem, particularly its challenging CK-graph family and modifications, as illustrated in Fig.~\ref{fig:CK_CK_like_schematic}. Following a simple, shallow QC build through local non-unitary gates, we obtain non-Hermitian QAAs through the FK and HD constructions, which we compare with a Grover-search-based Hermitian QAA (HM QAA). In the absence of noise, both the HD and FK QAA perform effectively, surpassing the HM QAA, which fails to amplify the target solution within the same polynomial evolution time. Under perturbations, on the other hand, the FK QAA's success probability and stability deteriorate rapidly as the process round or problem size increases. As a result, it is the HD QAA that combines polynomial efficiency with pseudospectral robustness, as we have summarized in Table~\ref{tab:qaa_comparison}. Finally, we propose an implementation of the HD QAA based on coupled optical waveguides with auxiliary lossy channels. Overall, these results establish that pseudospectral stability, together with a real and gapped spectrum, constitutes a fundamental design principle for practical non-Hermitian adiabatic quantum computation, and that the HD QAA provides an effective framework satisfying these requirements. 

Importantly, we discuss an additional complexity regarding such QAA through QC mapping: its ground states are, in fact, degenerate. Although there is no holonomy or off-diagonal matrix element within the degenerate subspace in the ideal case \cite{Simon1983Holonomy, WilczekZee1984}, noise may induce state mixing or even quantum walk within the degenerate subspace - this is not protected by the finite gap $\Omega$, which protects only against excitations out of the ground-state manifold \cite{ChildsFarhiPreskill2002}. A detailed measure-theoretic analysis of the degeneracy is in the Appendix. Interestingly, while this remains a severe issue for conventional Hermitian QAAs, the non-Hermitian QAA discussed in this work generally exhibits strong robustness: the non-Hermitian mechanism is self-focusing and, upon repeated iterations, amplifies relevant solutions and suppresses competing branches, regardless of the initialization or subsequent mixing, as long as the states remain somewhat non-orthogonal to the target solution. Such contrast is also evident in Fig.~\ref{fig:mis_noisy_dynamics}: even in the presence of general perturbations and ground-state degeneracy, the non-Hermitian HD QAA remains highly accurate; in contrast, the Hermitian QAA based on the Grover algorithm suffers a notable decrease in its success probability under large perturbations.  In addition, the QAA architecture is compatible with fault-tolerant adiabatic stabilizer codes for active suppression of such potential mixing \cite{Jordan2006, Lidar2008, YoungSarovarBlumeKohout2013}. 

Furthermore, it should be clarified that this work has not yet provided an algorithm with an exponential speedup at the complexity-theory level, nor does it claim to solve an NP-complete problem in polynomial time. In this context, it is essential to distinguish the wall-clock time (the actual physical time elapsed for a single execution) of the non-Hermitian QAA in this work from its total computational complexity. For example, in the physical implementation scheme given in Sec.~\ref{subsec:Potential_realizations}, the Hamiltonian is decoupled into $N=2^n$ mutually independent configuration blocks, corresponding to $\Theta(2^n)$ parallel coupled-waveguide systems, each with a wall-clock time $\sim\Theta(n^3)$, totaling $\Theta(2^n n^3)$ complexity and not an exponential speedup. 

To circumvent this spatial overhead, transitioning to a quantum system offers a natural solution. A quantum many-body state in the $n$-qubit Hilbert space $\mathcal{H}=(\mathbb{C}^2)^{\otimes n}$ coherently encodes an exponentially large number of classical configurations, thereby avoiding the enormous overhead of physically replicating the $2^n$ channels in the classical optical scheme. However, if the required non-Hermitian dynamics are implemented using standard post-selection schemes, the system will still face exponential complexity. For instance, consider embedding the HD QAA into continuously monitored Lindblad dynamics \cite{DalibardCastinMolmer1992, Minganti2020HybridLiouvillian}, the density matrix $\hat{\rho}$ evolves according to the Lindblad master equation:
\begin{equation}
\frac{d\hat{\rho}}{dt} = -i[H_0(s), \hat{\rho}] + \sum_k \left( C_k(s) \hat{\rho} C_k^\dagger(s) - \frac{1}{2} \{C_k^\dagger(s) C_k(s), \hat{\rho}\} \right),
\end{equation}
where $H_0(s)$ is the Hermitian Hamiltonian of the system and $C_k(s)$ are the quantum jump operators for the dissipation channels, representing sudden and irreversible exchanges of information or energy between the system and the environment, such as the emission of a photon. Then, if continuous monitoring detects no quantum jumps, the evolution is entirely governed by the effective non-Hermitian Hamiltonian:
\begin{equation}
H_0(s) - \frac{i}{2} \sum_k C_k^\dagger(s) C_k(s), 
\end{equation}
which, by properly setting $H_0(s)$ and $C_k(s)$, can equal the target HD QAA Hamiltonian $H_l(s)$ up to an overall normalization. Once a quantum jump is detected, however, that specific run is explicitly marked as a failure and discarded. Inevitably, the probability of a ``no-jump" quantum trajectory decays exponentially with the process time $T$ and thus the problem size $n$. Consequently, even though a successful branch requires only a polynomial wall-clock time $\sim\Theta(n^3)$, the overall required number of (parallelizable) repetitions remains exponential. This approach does not achieve a true exponential speedup in terms of computational complexity. 

Nevertheless, such parallelization or postselection, despite its exponential cost, is highly compatible with simpler fault-tolerant quantum hardware, a potentially worthwhile tradeoff given the resources required: logical depth, T-count, magic-state consumption, the number of ancillary qubits, and decoding overhead often face more severe constraints than those encountered when simply repeating shallow circuits \cite{Preskill2018NISQ, BravyiKitaev2005Magic, Fowler2012SurfaceCode, CampbellTerhalVuillot2017,
GidneyFowler2019MagicFactories}. For instance, a non-Hermitian HD QAA with postselection implementation will exhibit two advantages. First, a Hermitian QAA needs to evolve the quantum state coherently for an exponentially long time, which can easily collapse due to the accumulation of decoherence, gate errors, crosstalk, measurement errors, etc., making it unrealistic under present hardware conditions; instead, the proposed realization of HD QAA resorts to exponentially many short coherent trajectories, which are highly parallelizable and straightforwardly discarded and restarted when failed without much overhead. Second, errors in conventional deep circuits may accumulate invisibly and eventually contaminate the output; instead, the flags in the proposed implementation (e.g., the quantum jumps detected via continuous monitoring as described earlier) are straightforward to identify and remove, protecting the validity of the outputs \cite{Aaronson2005PostBQP, Bocharov2015RUS, Silva2023FragmentedQITE}. Thus, an infeasible, exponentially deep quantum process is transformed into a shallow, parallelizable, resettable quantum repetition process equipped with failure flags, thereby enabling a potential practical realization under the limited error-correction capabilities of the noisy intermediate-scale quantum (NISQ) era. 

Looking ahead, Rydberg atom arrays may become an emergent, genuinely many-body platform for realizing and exploring the HD QAA, thereby achieving a fundamental speedup in total complexity. The Rydberg blockade interaction $V_{ij}n_i n_j$ can naturally impose an energy penalty directly corresponding to the independent-set constraint $x_i x_j=0$ in the MIS problem, making it an ideal platform for implementing such quantum optimization algorithms \cite{Saffman2010RydbergQI, Browaeys2020RydbergReview, WuYou2021RydbergReview, Pichler2018RydbergMIS, Ebadi2022RydbergMIS, ZhaoYouWilczekWu2025HamiltonianMIS}. In addition, benefiting from capabilities such as atomically reconfigurable geometric arrangements, single-site addressing, tunable Rabi frequencies and detuning, strong interactions, and single-shot readout, together with recent experimental progress in driven-dissipative systems, Rydberg arrays have demonstrated a powerful ability to go beyond the limitations of Hermiticity and realize effective and controllable non-Hermitian terms \cite{Begoc2025ControlledDissipation, WuYou2024DissipativeTimeCrystal, Chen2025CollectiveDissipationRydberg}. Therefore, by harnessing their intrinsic, hardware-native dissipation mechanisms, Rydberg atom arrays hold the promise of significantly mitigating the exponential overhead typically associated with non-Hermitian implementations and provide a natural and promising hardware entry point for many-body non-Hermitian QAAs, including HD QAA.

\begin{acknowledgments}
We acknowledge helpful discussions with Biao Wu and support from the Shanghai Municipal Science and Technology Project (Grant No.25LZ2601100) and the National Key R\&D Program of China (Grants No.2022YFA1403700 and No.2021YFA1401900). 
\end{acknowledgments}

\bibliographystyle{apsrev4-2}
\bibliography{references}

\end{document}